\documentclass[twocolumn,twoside,letterpaper]{IEEEtran}
\usepackage[pdftex]{graphicx}
\usepackage[cmex10]{amsmath}
\usepackage{cite}
\usepackage{mathtools}
\usepackage{url}
\usepackage{color}
\usepackage{soul}
\usepackage{lipsum}

\begin{document}
\title{Widely-Linear Digital Self-Interference Cancellation in Direct-Conversion Full-Duplex Transceiver}

\author{Dani~Korpi,~
				Lauri~Anttila,
        Ville~Syrj\"{a}l\"{a},
        and~Mikko~Valkama \vspace{-6mm}
\thanks{Manuscript received October 6, 2013; revised March 24, 2014; accepted May 27, 2014. The published version of this article is available at http://dx.doi.org/10.1109/JSAC.2014.2330093}%
\thanks{D. Korpi, L. Anttila, V. Syrj\"{a}l\"{a}, and M. Valkama, are with the Department
of Electronics and Communications Engineering, Tampere University of Technology, PO Box 692, FI-33101, Tampere, Finland, e-mail: dani.korpi@tut.fi.}
\thanks{The research work leading to these results was funded by the Academy of Finland (under the project \#259915 "In-band Full-Duplex MIMO Transmission: A Breakthrough to High-Speed Low-Latency Mobile Networks"), the Finnish Funding Agency for Technology and Innovation (Tekes, under the project "Full-Duplex Cognitive Radio"), the Linz Center of Mechatronics (LCM) in the framework of the Austrian COMET-K2 programme, and Emil Aaltonen Foundation.}%
\thanks{\textcopyright \hspace{1pt} 2014 IEEE. Personal use of this material is permitted. Permission from IEEE must be obtained for all other uses, in any current or future media, including reprinting/republishing this material for advertising or promotional purposes, creating new collective works, for resale or redistribution to servers or lists, or reuse of any copyrighted component of this work in other works.}
}

\markboth{IEEE JOURNAL ON SELECTED AREAS IN COMMUNICATIONS}%
{Korpi \lowercase{\textit{et al.}}: Widely-Linear Digital Self-Interference Cancellation in Direct-Conversion Full-Duplex Transceiver}

\maketitle

\begin{abstract}
This article addresses the modeling and cancellation of self-interference in full-duplex direct-conversion radio transceivers, operating under practical imperfect radio frequency (RF) components. Firstly, detailed self-interference signal modeling is carried out, taking into account the most important RF imperfections, namely transmitter power amplifier nonlinear distortion as well as transmitter and receiver IQ mixer amplitude and phase imbalances. The analysis shows that after realistic antenna isolation and RF cancellation, the dominant self-interference waveform at receiver digital baseband can be modeled through a widely-linear transformation of the original transmit data, opposed to classical purely linear models. Such widely-linear self-interference waveform is physically stemming from the transmitter and receiver IQ imaging, and cannot be efficiently suppressed by classical linear digital cancellation. Motivated by this, novel widely-linear digital self-interference cancellation processing is then proposed and formulated, combined with efficient parameter estimation methods. Extensive simulation results demonstrate that the proposed widely-linear cancellation processing clearly outperforms the existing linear solutions, hence enabling the use of practical low-cost RF front-ends utilizing IQ mixing in full-duplex transceivers.
\end{abstract}


\vspace{-4mm}
\section{Introduction}
\IEEEPARstart{F}{ull-duplex} radio communications with simultaneous transmission and reception at the same radio frequency (RF) carrier is one of the emerging novel paradigms to improve the efficiency and flexibility of RF spectrum use. Some of the recent seminal works in this field are for example \cite{Radunovic09,Choi10,Jain11,Sahai11,Duarte10,Day12,Duarte12,Duarte122}, to name a few.  Practical realization and implementation of small and low-cost full-duplex transceivers, e.g., for mobile cellular radio or local area connectivity devices, are, however, still subject to many challenges. One of the biggest problems is the so called self-interference (SI), which is stemming from the simultaneous transmission and reception at single frequency, thus causing the strong transmit signal to couple directly to the receiver path. 

In general, the transmitter and receiver may use either the same \cite{Knox12,Cox13,Phungamngern13} or separate but closely-spaced antennas \cite{Choi10,Jain11,Sahai11,Duarte10}. In this work, we focus on the case of separate antennas where depending on, e.g., the deployed physical antenna separation and transmit power, the level of the coupling SI signal can be in the order of 60--100 dB stronger than the actual received signal of interest at receiver input, especially when operating close to the sensitivity level of the receiver chain. To suppress such SI inside the transceiver, various antenna-based solutions \cite{Choi10,Everett13,Aryafar12,Sahai11}, active analog/RF cancellation methods \cite{Choi10,Jain11,Bharadia13,Sahai11,McMichael12,Lee13,Duarte122} and digital baseband cancellation techniques \cite{Anttila13,Bharadia13,Ahmed13,Jain11} have been proposed in the literature. In addition, a general analysis about the overall performance of different linear SI cancellation methods is performed in \cite{Duarte12}, while in \cite{Riihonen13}, SI cancellation based on spatial-domain suppression is compared to subtractive time-domain cancellation, and rate regions are calculated for the two methods.

However, the performance of the SI cancellation mechanisms based on linear processing is usually limited by the analog/RF circuit non-idealities occurring within the full-duplex transceiver. For this reason, some of the most prominent types of such analog/RF circuit non-idealities have been analyzed in several recent studies. The phase noise of the transmitter and receiver oscillators has been analyzed in \cite{Riihonen1222,Sahai12,Syrjala13}. It was observed that the phase noise can significantly limit the amount of achievable SI suppression, especially when using two separate oscillators for transmitter and receiver. A signal model including the effect of phase noise is also investigated in \cite{Sahai122}, where the feasibility of asynchronous full-duplex communications is studied. In \cite{Ahmed133}, the effects of the receiver chain noise figure and the quantization noise are also taken into account, in addition to phase noise. The authors then provide an approximation for the rate gain region of a full-duplex transceiver under the analyzed impairments. The effect of quantization noise is also analyzed in \cite{Riihonen122}, where the relation between analog and digital cancellation under limited dynamic range for the analog-to-digital converter is studied. The existence of several non-idealities in the transmit chain, including power amplifier (PA) nonlinearity, is acknowledged in \cite{Li11}. As a solution, the authors propose taking the cancellation signal from the output of the transmitter chain to exclude these non-idealities from the signal path. In \cite{Bliss12}, several non-idealities in the transmitter and receiver chains are also considered, including nonlinearities and IQ mismatch. Their effect is then studied in terms of the achievable SI suppression with linear processing and received spatial covariance eigenvalue distribution. In \cite{Zheng13}, the transmitter non-idealities are modelled as white noise, the power of which is dependent on the transmit power, and their effect is analyzed in the context of cognitive radios. In \cite{Korpi13}, comprehensive distortion calculations, taking into account several sources of nonlinear distortion, among other things, are reported. The findings indicate that, especially with high transmit powers, the nonlinear distortion of the transmitter PA and receiver front-end components can significantly contribute to the SI waveform and thus hinder the efficiency of purely linear digital cancellation. Stemming from the findings of these studies, novel nonlinear cancellation processing solutions have been recently proposed in \cite{Anttila13,Ahmed13,Bharadia13} to suppress such nonlinear SI, in addition to plain linear SI, at receiver digital baseband.

In addition to phase noise and nonlinearities, also other RF imperfections can impact the SI waveform and its cancellation. One particularly important imperfection in IQ processing based architectures is the so-called IQ imbalance and the corresponding inband IQ image or mirror component \cite{Anttila11}. On the transmitter side, in general, such IQ imbalance is contributing to the transmitter error vector magnitude, and possibly also to adjacent channel leakage and spurious emissions, depending on the transmit architecture. As a practical example, 3GPP Long Term Evolution (LTE)/LTE-Advanced radio system specifications limit the minimum attenuation for the inband image component in mobile user equipment transmitters to 25 dB or 28 dB, depending on the specification release \cite{LTE_specs}. Such image attenuation is sufficient in the transmission path, but when considering the full-duplex device self-interference problem, the IQ image of the SI signal is additional interference leaking to the receiver path. Furthermore, additional IQ imaging of the SI signal takes place in the receiver path. In \cite{Hua12}, the effect of IQ imbalance on a full-duplex transceiver was noticed in the measurements, as it was causing clear residual SI after all the cancellation stages. However, there is no previous work on compensating the SI mirror component caused by the IQ imbalance of the transmitter and receiver chains. Overall, there is relatively little discussion in the existing literature about the effect of IQ imbalance on the performance of a full-duplex transceiver. In part, this is due to the high-cost high-quality equipment that has been used to demonstrate the full-duplex transceiver principle in the existing implementations. For example, the WARP platform, which has been used at least in \cite{Choi10,Jain11,Duarte10,Bharadia13,Duarte12}, provides an attenuation in the excess of 40--50 dB for the inband image component \cite{maxim_datasheet}. As our analysis in this article indicates, this is adequate to decrease the mirror component sufficiently low for it to have no significant effect on the performance of the full-duplex transceiver. Furthermore, if properly calibrated high-end laboratory equipment is used, the image attenuation can easily be even in the order of 60--80~dB, meaning that the effect of IQ imbalance is totally negligible. \textit{However, for a typical mobile transceiver with low-cost mass-market analog/RF components, the image attenuation is generally significantly less, as already mentioned} \cite{LTE_specs}. This means that also the effect of IQ imbalance in a full-duplex transceiver must be analyzed and taken into account.

In this article, we firstly carry out detailed modeling for the SI waveform in different stages of the transmitter-receiver coupling path, taking into account the effects of transmitter and receiver IQ image components, as well as transmitter PA nonlinearities. Incorporating then also the effects of realistic multipath antenna coupling, linear analog/RF cancellation and linear digital baseband cancellation, the powers of the remaining SI components at the output of the full coupling and processing chain are analyzed. This analysis shows that, with realistic component values and linear cancellation processing, the IQ image of the classical linear SI is heavily limiting the receiver path signal-to-noise-plus-interference ratio (SINR). Such observation has not been reported earlier in the literature. Motived by these findings, a novel widely-linear (WL) digital SI canceller is then developed, where not only the original transmit data, but also its complex conjugate, modeling the IQ imaging, are processed to form an estimate of the SI signal. Efficient parameter estimation methods are also developed to estimate the cancellation parameters of the proposed WL structure through WL least-squares model fitting. The proposed WL SI canceller is shown by analysis, and through extensive simulations, to substantially improve the SI cancellation performance in the presence of practical IQ imaging levels, compared to classical purely linear processing, and it can hence enable full-duplex transceiver operation with realistically IQ balanced low-cost user equipment RF components.

The rest of this article is organized as follows. In Section~\ref{sec:signal_model}, the structure of the considered full-duplex transceiver and its baseband-equivalent model are presented, alongside with the overall self-interference signal model and its simplified version. Also principal system calculations, in terms of the powers of the different self-interference terms, are carried out. The proposed method for widely-linear digital cancellation and the estimation procedure for the coefficients are then presented in Section~\ref{sec:wl_canc}. In Section~\ref{sec:simulations}, the performance of widely-linear digital cancellation under different scenarios is analyzed with full waveform simulations. Finally, the conclusions are drawn in Section~\ref{sec:conc}.

\section{Full-Duplex Transceiver and Self-Interference}
\label{sec:signal_model}

\begin{figure*}[!t]
\centering
\includegraphics[width=0.85\textwidth]{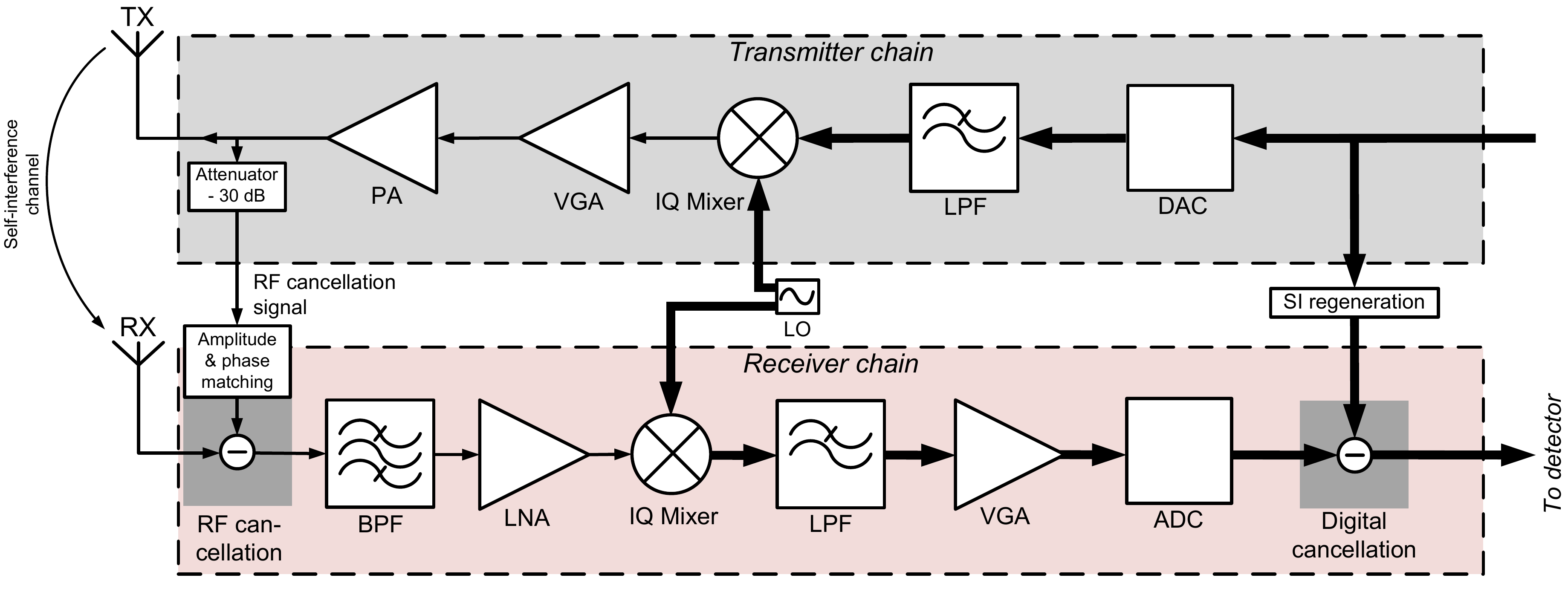}
\caption{A block diagram of the assumed direct-conversion full-duplex transceiver.}
\label{fig:block_diagram}
\end{figure*}

The structure of the analyzed full-duplex transceiver is presented in Fig.~\ref{fig:block_diagram}. It can be observed that the transceiver follows a typical direct-conversion architecture \cite{Gu06,Parssinen99,Yoshida03}, which is well-known in previous literature, and thus it is not discussed here in detail. This architecture is chosen due to its simple structure and wide applications in modern wireless transceivers. The actual IQ imaging problem is caused by the IQ mixers at both the transmitter and receiver chains. Due to the inherent mismatches between the amplitudes and phases of the I- and Q-branches, the mirror image of the original signal is added on top of it, with certain image attenuation \cite{Anttila11}. In this paper, we assume that the level of this image attenuation is similar to what is specified in 3GPP LTE specifications \cite{LTE_specs}, namely 25 dB.

The actual analysis of the full-duplex transceiver and SI waveform at different stages of the transceiver is done next by using baseband-equivalent models. The block diagram of the overall baseband-equivalent model is shown in Fig.~\ref{fig:baseband_model}, alongside with the principal mathematical or behavioral model for each component, propagation of the transmitted signal, and the corresponding variable names. In the following subsection, a complete characterization for the effective SI waveform in different stages of the transceiver is provided, using the same notations as in Fig.~\ref{fig:baseband_model}, taking into account transmitter IQ imaging and PA distortion, realistic multipath coupling channel, realistic linear analog/RF cancellation, receiver IQ imaging, and receiver linear cancellation. Stemming from this, the powers of the different SI terms at receiver digital baseband are then analysed in Subsection~\ref{sec:syscalc}.

\subsection{Self-Interference Signal Model with Practical RF Components}

\begin{figure*}[!t]
\centering
\includegraphics[width=0.85\textwidth]{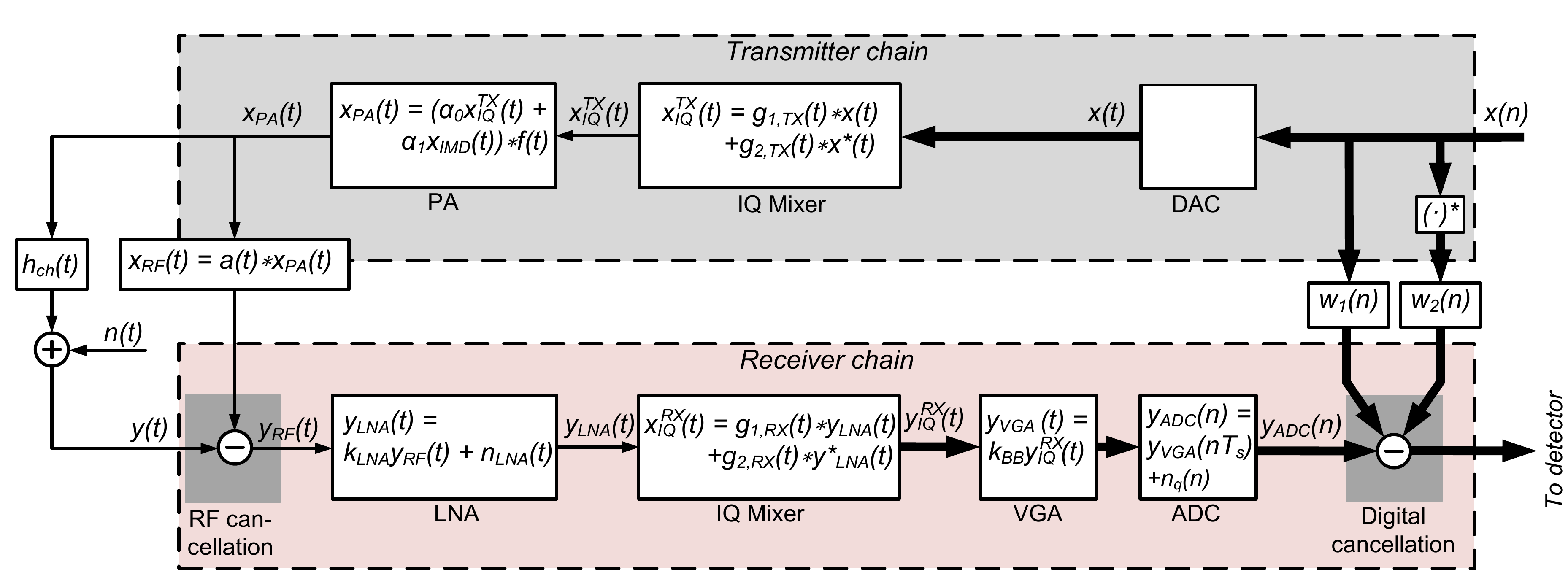}
\caption{A baseband-equivalent model of the analyzed full-duplex transceiver, alongside with the signals propagating at the different stages of the transceiver. Digital baseband SI cancellation is already showing, at structural level, the proposed WL SI cancellation where both the original transmit data and its complex conjugate are processed.}
\label{fig:baseband_model}
\end{figure*}

The complex baseband transmitted signal is denoted by $x(n)$, or by $x(t)$ after digital-to-analog conversion. It is assumed that the power of $x(t)$ is such that the desired transmit power is reached after the amplification by IQ mixer and PA, i.e., the transmitter VGA is omitted from the baseband-equivalent model. This is done to make the notation simpler and thus more illustrative. In addition, the signal $x(n)$ is perfectly known, as it is generated within the transceiver. In this analysis, the digital-to-analog converters (DACs) and low-pass filters (LPFs) are assumed to be ideal, but the IQ mixer is assumed to have some imbalance between the I and Q branches. The signal at the output of the transmit IQ mixer can be now written as
\begin{align}
	x^{TX}_{IQ}(t) &= g_{1,TX}(t) \star x(t) + g_{2,TX}(t) \star x^{\ast}(t) \text{,} \label{eq:x_tx}
\end{align}
where $g_{1,TX}(t)$ is the response for the direct signal component, and $g_{2,TX}(t)$ is the response for the image component \cite{Anttila11}. Here $(\cdot)^{\ast}$ indicates the complex conjugate and $\star$ denotes the convolution operation. Above-kind of transformations, where both direct and complex-conjugated signals are filtered and finally summed together, are typically called widely-linear in the literature, see e.g. \cite{Schreier10,Picinbono95}.

The quality of the IQ mixer can be quantified by the image rejection ratio (IRR). With the variables used in this analysis, it can be defined for the transmitter as
\begin{align}
	\mathit{IRR}_{TX}(f) = \frac{|G_{1,TX}(f)|^2}{|G_{2,TX}(f)|^2} \text{,} \label{eq:irr}
\end{align}
where $G_{1,TX}(f)$ and $G_{2,TX}(f)$ are the frequency-domain representations of $g_{1,TX}(t)$ and $g_{2,TX}(t)$, respectively \cite{Anttila11}. A similar characterization can obviously be established also for the receiver IQ mixer, referred to as $\mathit{IRR_{RX}}(f)$.

Before transmission, the signal is amplified with a nonlinear PA. In this analysis, we model the PA response with a Hammerstein nonlinearity \cite{Ding04,Morgan06} given as
\begin{align}
	x_{PA}(t) &= \left(\alpha_{0} x^{TX}_{IQ}(t) + \alpha_{1} x^{TX}_{IQ}(t) |x^{TX}_{IQ}(t)|^2\right) \star f(t) \text{,} \label{eq:x_pa_long}
\end{align}
where $\alpha_{0}$ is the linear gain, $\alpha_{1}$ is the gain of the third order component, and $f(t)$ is the memory model of the PA. For simplicity, we write $x_{IMD}(t) = x^{TX}_{IQ}(t) |x^{TX}_{IQ}(t)|^2$, and use this to refer to the third order nonlinear component. Thus, we can rewrite \eqref{eq:x_pa_long} as
\begin{align}
	x_{PA}(t) &= \left(\alpha_{0} x^{TX}_{IQ}(t) + \alpha_{1} x_{IMD}(t)\right) \star f(t) \text{.} \label{eq:x_pa}
\end{align}
It is obvious that true PAs contain also distortion components beyond third-order, but in this analysis we make a simplification and focus only on the third-order distortion, as that is in practice always the strongest nonlinearity at PA output.

The transmitted signal is next coupled back to the receive antenna, thus producing SI. In this analysis, to simplify the notations, it is assumed that there is no actual received signal of interest. This will decrease the complexity of the equations while having no significant effect on the results, as the purpose is to characterize the final SI waveform at receiver baseband. Thus, the signal at the input of the receiver chain is of the following form:
\begin{align}
	y(t) &= h_{ch}(t) \star x_{PA}(t) + n_{th}(t) \text{,} \label{eq:y}
\end{align}
where $h_{ch}(t)$ is the multipath coupling channel between transmit and receive antennas, and $n_{th}(t)$ denotes thermal noise. To suppress the SI before it enters the LNA, RF cancellation is performed. The signal after RF cancellation can be expressed as
\begin{align}
	y_{RF}(t) &= y(t) - a(t) \star x_{PA}(t) \text{,} \label{eq:y_rf}
\end{align}
where $a(t)$ is typically an estimate for the main path of the coupling channel \cite{Duarte12,Choi10,Jain11}. In other words, $a(t)$ is a one tap filter, depicting the delay, phase, and attenuation of the main coupling propagation. Thus, in this analysis it is assumed that RF cancellation attenuates only the direct coupling component. Furthermore, in the numerical experiments and results, $a(t)$ is chosen so that it provides the desired amount of SI attenuation at RF, e.g., 30 dB, as has been reported in practical experiments \cite{Duarte10,Duarte12,McMichael12}. This is done by tuning the error of its attenuation and delay, modeling realistic RF cancellation.

Next, the received signal is amplified by the low-noise amplifier (LNA). The output signal of the LNA can be written as
\begin{align}
	y_{LNA}(t) &= k_{LNA} y_{RF}(t) + n_{LNA}(t) \text{,} \label{eq:y_lna}
\end{align}
where $k_{LNA}$ is the complex gain of the LNA, and $n_{LNA}(t)$ is the increase in the noise floor caused by the LNA. The nonlinearity effects of the receiver chain amplifiers are not taken into account in this paper, as they are insignificant in comparison to the other distortion components under typical circumstances \cite{Korpi13}. This will also significantly simplify the analysis.

Similar to the transmitter, the receiver IQ mixer has IQ imbalance, and thus it produces an image component of the total signal entering the mixer, including the SI. The signal at the output of the receiver IQ mixer can be now expressed as
\begin{align}
y^{RX}_{IQ}(t) &= g_{1,RX}(t) \star y_{LNA}(t) + g_{2,RX}(t) \star y_{LNA}^{\ast}(t) \text{,} \label{eq:y_rx}
\end{align}
where $g_{1,RX}(t)$ is the response for the direct signal component, and $g_{2,RX}(t)$ is the response for the receiver image component.

Finally, the signal is amplified by the variable gain amplifier (VGA) to match its waveform dynamics to the voltage range of the analog-to-digital converter (ADC) and then digitized. The digitized signal can be written as
\begin{align}
	y_{ADC}(nT_s) &= k_{BB} y^{RX}_{IQ}(nT_s) + n_{q}(nT_s) \text{,} \label{eq:y_adc}
\end{align}
where $k_{BB}$ is the complex baseband gain of the VGA, $T_s$ is the sampling time, and $n_q(nT_s)$ denotes quantization noise. In the continuation, we drop the sampling interval $T_s$ from the equations for brevity and use only the discrete-time index $n$.

To express the residual SI signal at the digital domain in terms of the known transmit data $x(n)$, a complete equation for $y_{ADC}(n)$ is next derived by substituting \eqref{eq:x_tx} to \eqref{eq:x_pa}, \eqref{eq:x_pa} to \eqref{eq:y} and so on. After these elementary manipulations, we arrive at the following equation for $y_{ADC}(n)$, with respect to $x(n)$ and fundamental system responses:
\begin{align}
	y_{ADC}(n) &= h_{1}(n) \star x(n) + h_{2}(n) \star x^{\ast}(n)\nonumber\\
	&+ h_{IMD}(n) \star x_{IMD}(n) + h_{IMD,im}(n) \star x^{\ast}_{IMD}(n) \nonumber\\
	&+ h_{n}(n) \star n_{tot}(n) + h_{n,im}(n) \star n^{\ast}_{tot}(n)+ n_q(n) \text{,} \label{eq:y_c_compact}
\end{align}
where we have defined a total noise signal as $n_{tot}(n) = k_{LNA} n_{th}(n) + n_{LNA}(n)$, including the thermal noise at the input of the receiver chain and the additional noise produced by the LNA. The channel responses of the individual signal components can be written as follows:
\begin{align}
	h_{1}(n) &= k_{BB} k_{LNA} \alpha_{0} g_{1,TX}(n) \star g_{1,RX}(n) \star f(n)\nonumber\\
	&\star (h_{ch}(n) - a(n)) +k_{BB} k^{\ast}_{LNA} \alpha^{\ast}_{0}g^{\ast}_{2,TX}(n)\nonumber\\
	& \star g_{2,RX}(n) \star f^{\ast}(n) \star (h_{ch}^{\ast}(n) - a^{\ast}(n)) \label{eq:h_1}\\[2mm]
	h_{2}(n) &= k_{BB} k_{LNA} \alpha_{0} g_{2,TX}(n) \star g_{1,RX}(n) \star f(n)\nonumber\\
	&\star (h_{ch}(n) - a(n)) +k_{BB} k^{\ast}_{LNA} \alpha^{\ast}_{0} g^{\ast}_{1,TX}(n) \nonumber\\
	&\star g_{2,RX}(n) \star f^{\ast}(n) \star (h_{ch}^{\ast}(n) - a^{\ast}(n)) \label{eq:h_2}\\[2mm]
	h_{IMD}(n) &= k_{BB} k_{LNA} \alpha_{1} g_{1,RX}(n) \star f(n) \nonumber\\
	&\star (h_{ch}(n) - a(n)) \label{eq:h_x_imd} \\[2mm]
	h_{IMD,im}(n) &= k_{BB} k^{\ast}_{LNA} \alpha^{\ast}_{1} g_{2,RX}(n) \star f^{\ast}(n) \nonumber\\
	&\star (h_{ch}^{\ast}(n) - a^{\ast}(n)) \label{eq:h_x_imd_conj} \\[2mm]
	h_{n}(n) &= k_{BB} g_{1,RX}(n) \label{eq:n} \\[2mm]
	h_{n,im}(n) &= k_{BB} g_{2,RX}(n) \text{.} \label{eq:n_conj}
\end{align}
As can be seen in \eqref{eq:y_c_compact}, the total SI at receiver digital baseband contains not only the linear SI but also its complex conjugate. These different components of the SI signal are hereinafter referred to as linear SI and conjugate SI, respectively. In addition to these signal components, PA-induced IMD and its complex-conjugate, which will similarly be referred to as IMD and conjugate IMD, are also present in the total SI signal.

Using the above equations, it is possible to describe the effect of conventional linear digital SI cancellation, which can attenuate only the linear SI component. Corresponding to the notation in Fig.~\ref{fig:baseband_model}, where the linear channel estimate is denoted by $w_1(n)$, the signal after linear digital cancellation can be expressed as
\begin{align}
	y_{LDC}(n) &= y_{ADC}(n) - w_1(n) \star x(n) \nonumber\\
	&= (h_{1}(n)- w_1(n)) \star x(n) + h_{2}(n) \star x^{\ast}(n)\nonumber\\
	&+ h_{IMD}(n) \star x_{IMD}(n) + h_{IMD,im}(n) \star x^{\ast}_{IMD}(n)\nonumber\\
	&+ h_{n}(n) \star n_{tot}(n) + h_{n,im}(n) \star n^{\ast}_{tot}(n) + n_q(n) \text{.} \label{eq:y_c_dc}
\end{align}
From \eqref{eq:y_c_dc} it can be observed that $w_1(n)$ should estimate the channel of the linear SI component, i.e., $w_1(n) = \hat{h}_1(n)$. However, even with a perfect estimate of $h_1(n)$, the signal $y_{LDC}(n)$ can still be substantial interference from weak desired signal perspective. This is mainly due to the conjugate SI, IMD, and conjugate IMD. Obviously, there is also some thermal noise, but typically it is not significantly limiting the performance of a full-duplex transceiver.

Notice that nonlinear distortion has been shown earlier in the literature to limit the achievable SINR of a full-duplex radio, and there are also methods for attenuating it \cite{Anttila13,Ahmed13,Bharadia13}. However, there is no previous work on analyzing and attenuating the conjugate SI signal, which is relative to $x^{\ast}(n)$ in our notations. In the next subsection, we will analyze the relative strength of this conjugate SI through principal power calculations, and show that with typical RF component specifications, it is the dominant SINR limiting phenomenon. In Section~\ref{sec:wl_canc}, we then also provide a method for suppressing the conjugate SI in the digital domain by processing the original transmit data in a widely-linear manner with two filters $w_1(n)$ and $w_2(n)$, marked also in Fig.~\ref{fig:baseband_model}. However, in the following subsection it is still first assumed that $w_2(n) = 0$, and thus no compensation is done for the conjugate SI, in order to properly quantify and illustrate the limitations of classical linear SI cancellation.

\subsection{Principal System Calculations for Different Distortion Terms}
\label{sec:syscalc}

In order to analyze and illustrate the relative levels of the different distortion components of the overall SI signal, a somewhat simplified scenario is first considered. More specifically, the frequency-dependent characteristics of different distortion components are neglected, which allows for the equations to be presented in a more compact and illustrative form. Furthermore, as we are here primarily interested in the average powers of different distortion components, neglecting inband frequency-dependency is well justified. If we now denote the general impulse function by $\delta(n)$, the different system impulse responses of the distortion components can be expressed as follows: $g_{1,TX}(n) \approx g_{1,TX} \delta(n)$, $g_{2,TX}(n) \approx g_{2,TX} \delta(n)$, $g_{1,RX}(n) \approx g_{1,RX} \delta(n)$, $g_{2,RX}(n) \approx g_{2,RX} \delta(n)$, $f(n) \approx \delta(n)$, $h_{ch}(n) \approx h_{ch} \delta(n)$, $a(n) \approx a \delta(n)$, and $w_1(n) \approx w_1 \delta(n)$. By substituting these simplified terms into \eqref{eq:h_1}--\eqref{eq:n_conj}, we can rewrite them as
\begin{align}
	h_{1}(n) &\approx k_{BB} k_{LNA} \alpha_{0} g_{1,TX} g_{1,RX} (h_{ch} - a) \delta(n)\nonumber\\
	 &+k_{BB} k^{\ast}_{LNA} \alpha^{\ast}_{0} g^{\ast}_{2,TX} g_{2,RX} (h_{ch}^{\ast} - a^{\ast}) \delta(n) \label{eq:h_1_simp}\\
	h_{2}(n) &\approx k_{BB} k_{LNA} \alpha_{0} g_{2,TX} g_{1,RX}(h_{ch} - a) \delta(n)\nonumber\\
	 &+ k_{BB} k^{\ast}_{LNA} \alpha^{\ast}_{0} g^{\ast}_{1,TX} g_{2,RX} (h_{ch}^{\ast} - a^{\ast}) \delta(n) \label{eq:h_2_simp}\\
  h_{IMD}(n) &\approx k_{BB} k_{LNA} \alpha_{1} g_{1,RX} (h_{ch} - a) \delta(n) \label{eq:h_x_imd_simp} \\
	h_{IMD,im}(n) &\approx k_{BB} k^{\ast}_{LNA} \alpha^{\ast}_{1} g_{2,RX} (h_{ch}^{\ast} - a^{\ast}) \delta(n) \label{eq:h_x_imd_conj_simp} \\
	h_{n}(n) &\approx k_{BB} g_{1,RX} \delta(n) \label{eq:n_simp} \\
	h_{n,im}(n) &\approx k_{BB} g_{2,RX} \delta(n) \text{.} \label{eq:n_conj_simp}
\end{align}
Furthermore, as the magnitude of the term $g^{\ast}_{2,TX} g_{2,RX}$ is very small in comparison to the magnitude of $g_{1,TX} g_{1,RX}$, even when considering a relatively low image attenuation, we can write \eqref{eq:h_1_simp} as
\begin{align}
h_{1}(n) &\approx k_{BB} k_{LNA} \alpha_{0} g_{1,TX} g_{1,RX} (h_{ch} - a) \delta(n) \label{eq:h_1_simp2}
\end{align}
Using \eqref{eq:h_2_simp}--\eqref{eq:h_1_simp2}, $y_{LDC}(n)$ can then be expressed in a simplified form as follows:
\begin{align}
	y_{LDC}(n) &\approx (h_1 - w_1) x(n) + h_{2} x^\ast (n) + h_{IMD} x_{IMD}(n)\nonumber\\
	&+ h_{IMD,im} x^{\ast}_{IMD}(n) + h_{n} n_{tot}(n)+ h_{n,im}n^{\ast}_{tot}(n)\nonumber\\
	&+n_q(n) \nonumber\\
	&= (k_{BB} k_{LNA} \alpha_{0} g_{1,TX} g_{1,RX} (h_{ch} - a) - w_1) x(n) \nonumber\\
	&+(k_{BB} k_{LNA} \alpha_{0} g_{2,TX} g_{1,RX}(h_{ch} - a) \nonumber\\
	&+ k_{BB} k^{\ast}_{LNA} \alpha^{\ast}_{0} g^{\ast}_{1,TX} g_{2,RX} (h_{ch}^{\ast} - a^{\ast})) x^\ast (n) \nonumber\\
	&+ k_{BB} k_{LNA} \alpha_{1} g_{1,RX} (h_{ch} - a) x_{IMD}(n) \nonumber\\
	&+ k_{BB} k^{\ast}_{LNA} \alpha^{\ast}_{1} g_{2,RX} (h^{\ast}_{ch} - a^{\ast}) x^{\ast}_{IMD}(n) \nonumber\\
	&+ k_{BB} g_{1,RX} n_{tot}(n) + k_{BB} g_{2,RX} n^{\ast}_{tot}(n)+n_q(n) \text{.} \label{eq:y_ldc_simp}
\end{align}

From \eqref{eq:y_ldc_simp}, it is now possible to calculate the powers of the different signal components. These powers are defined as follows:
{\small
\begin{align}
	p_{SI} &= E[|k_{BB} k_{LNA} \alpha_{0} g_{1,TX} g_{1,RX}(h_{ch} - a)- w_1|^2 |x(n)|^2]\nonumber\\
	&= E[|k_{BB} k_{LNA} \alpha_{0} g_{1,TX} g_{1,RX}(h_{ch} - a)- w_1|^2]\nonumber\\
	&\times E[|x(n)|^2]\nonumber\\
	&= E[|k_{BB} k_{LNA}\alpha_{0} g_{1,TX} g_{1,RX} (h_{ch} - a)- w_1|^2] p_{x} \label{eq:p_si} \\
	p_{SI,im} &= E[|k_{BB} k_{LNA} \alpha_{0} g_{2,TX} g_{1,RX}(h_{ch} - a) \nonumber\\
	&+ k_{BB} k^{\ast}_{LNA} \alpha^{\ast}_{0} g^{\ast}_{1,TX} g_{2,RX} (h_{ch}^{\ast} - a^{\ast})|^2] p_{x} \label{eq:p_si_im} \\[2mm]
	p_{IMD} &= E[|k_{BB} k_{LNA} \alpha_{1} g_{1,RX} (h_{ch} - a)|^2] p_{x,IMD} \label{eq:p_imd} \\[2mm]
	p_{IMD,im} &= E[|k_{BB} k^{\ast}_{LNA} \alpha^{\ast}_{1} g_{2,RX} (h^{\ast}_{ch} - a^{\ast})|^2] p_{x,IMD}  \label{eq:p_imd_im}\\
	p_{noise} &= E[|k_{BB}  g_{1,RX}|^2] p_{n} \label{eq:p_n}\\[2mm]
	p_{noise,im} &= E[|k_{BB}  g_{2,RX}|^2] p_{n} \label{eq:p_n_im}\\[2mm]
	p_{q} &= E[|n_q(n)|^2] = \frac{p_{AD}}{\mathit{snr}_{ADC}} \text{,}\label{eq:p_q}
\end{align}}%
where $E[\cdot]$ denotes the expected value, $p_{x} = E[|x(n)|^2]$ is the power of the signals $x(n)$ and $x^{\ast}(n)$, as conjugation does not affect the power of the signal, $p_{x,IMD} = E[|x_{IMD}(n)|^2]$ is the power of the signals $x_{IMD}(n)$ and $x^\ast_{IMD}(n)$, $p_{n} = E[|n_{tot}(n)|^2]$ is the power of the signals $n_{tot}(n)$ and $n^{\ast}_{tot}(n)$, and $p_{q} = E[|n_q(n)|^2]$ is the power of quantization noise. Here, $p_q$ is defined in terms of the total signal power at the input of the ADC, $p_{AD}$, and the SNR of the ADC, $\mathit{snr}_{ADC} = 10^{(6.02b+4.76-\mathit{PAPR})/10}$, where $b$ is the number of bits at the ADC, and $\mathit{PAPR}$ is the peak-to-average-power ratio in dB \cite{Gu06}.

For further simplicity, the power of the total noise term $p_{n} = E[|n_{tot}(n)|^2]$ can be expressed in a more compact manner. As the LNA constitutes for most of the noise factor of the receiver chain, denoted by $F$, it can be appoximated with very little error that LNA has a noise factor of $F$, and no additional noise is produced throughout the rest of the receiver chain. Thus, if the power of the thermal noise at the input of the receiver chain is defined as $p_{th} = E[|n_{th}(n)|^2]$, we can write $p_n = E[|n_{tot}(n)|^2] = E[|k_{LNA} n_{th}(n) + n_{LNA}(n) |^2] = |k_{LNA}|^2 F p_{th}$, based on the definition of the noise factor \cite{Gu06}.

In addition, to express the power levels using the defined parameters, the amount of achieved RF cancellation with respect to the error between $h_{ch}$ and $a$ must be defined. This can be done by comparing the power of the total SI signal before and after RF cancellation. Based on \eqref{eq:y}, the SI signal before RF cancellation is $y_{SI}(t) = h_{ch}(t) \star x_{PA}(t)$. After RF cancellation, the SI signal is $y_{RF,SI}(t) = y_{SI}(t) - a(t)\star x_{PA}(t) = (h_{ch}(t) - a(t)) \star x_{PA}(t) $, based on \eqref{eq:y_rf}. Taking the previously defined simplifications into account, we can write $y_{SI}(t) \approx h_{ch} x_{PA}(t)$ and $y_{RF,SI}(t) = (h_{ch} - a) x_{PA}(t)$. Thus, the amount of RF cancellation can be defined as
\begin{align}
	&|a_{RF}|^2 = \frac{E[|y_{RF,SI}(t)|^2]}{E[|y_{SI}(t)|^2]} = \frac{E[|(h_{ch}-a) x_{PA}(t)|^2]}{E[|h_{ch} x_{PA}(t)|^2]}\nonumber\\
	&= \frac{E[|h_{ch}-a|^2] E[|x_{PA}(t)|^2]}{E[|h_{ch}|^2] E[|x_{PA}(t)|^2]} = \frac{E[|h_{ch}-a|^2]}{E[|h_{ch}|^2]} \text{.} \label{eq:a_rf}
\end{align}
Here it is assumed that the instantaneous path loss of the SI coupling channel, denoted by $|h_{ch}|^2$, and the error of the RF cancellation channel estimate, denoted by $(h_{ch}-a)$, are circular and normally distributed. The amount of antenna attenuation is now defined as $|a_{ant}|^2 = E[|h_{ch}|^2]$. Using \eqref{eq:a_rf}, it is then possible to define the power of the error between the SI coupling channel and the channel estimate for RF cancellation as
\begin{align}
	E[|h_{ch}-a|^2] = E[|h_{ch}|^2] |a_{RF}|^2 =|a_{ant}|^2 |a_{RF}|^2  \text{.}  \label{eq:rf_error}
\end{align}

The amount of achieved linear digital cancellation is defined next as the decrease in the power of the linear SI component $x(n)$. Before digital cancellation, the linear SI signal can be expressed as $y_{ADC,SI}(n) = h_1(n)\star x(n) \approx h_1 x(n)$, and after linear digital cancellation as $y_{LDC,SI}(n) = y_{ADC,SI}(n)- w_1(n)\star x(n) \approx (h_1 - w_1) x(n)$. The attenuation of the linear SI power by linear digital cancellation can then be expressed as follows:
\begin{align}
	&|a_{LDC}|^2 = \frac{E[|y_{LDC,SI}(n)|^2]}{E[|y_{ADC,SI}(n)|^2]} = \frac{E[|(h_1 - w_1) x(n)|^2]}{E[|h_1 x(n)|^2]}\nonumber\\
	&= \frac{E[|h_1 - w_1|^2] E[|x(n)|^2]}{E[|h_1|^2] E[|x(n)|^2]} = \frac{E[|h_1-w_1|^2]}{E[|h_1|^2]} \text{.} \label{eq:a_ldc}
\end{align}
Now, by using \eqref{eq:a_ldc}, it is possible to express $E[|h_1-w_1|^2]$ in terms of $|a_{LDC}|^2$ as follows:
\begin{align}
	E[|h_1-w_1|^2] = |a_{LDC}|^2 E[|h_1|^2] \text{,} \label{eq:digcanc}
\end{align}
where $h_1 = k_{BB} k_{LNA} \alpha_{0} g_{1,TX} g_{1,RX} (h_{ch} - a)$.

Now, by substituting \eqref{eq:rf_error} to \eqref{eq:p_si}--\eqref{eq:p_imd_im}, \eqref{eq:digcanc} to \eqref{eq:p_si}, and $p_n = |k_{LNA}|^2 F p_{th}$ to \eqref{eq:p_n}--\eqref{eq:p_n_im}, we can finally express the power levels of all the different signal components as follows:
\begin{align}
	p_{SI} &= |a_{LDC}|^2 E[|k_{BB}|^2 |k_{LNA}|^2 |\alpha_{0}|^2 |h_{ch} - a|^2\nonumber\\
	&\times|g_{1,TX}|^2 |g_{1,RX}|^2]  p_{x} \nonumber\\
	&= |a_{LDC}|^2 |k_{BB}|^2 |k_{LNA}|^2 |\alpha_{0}|^2 E[|h_{ch} - a|^2]\nonumber\\
	&\times |g_{1,TX}|^2 |g_{1,RX}|^2  p_{x} \nonumber\\
	&= |a_{LDC}|^2 |k_{BB}|^2 |k_{LNA}|^2 |\alpha_{0}|^2 |a_{ant}|^2 |a_{RF}|^2\nonumber\\
	&\times |g_{1,TX}|^2 |g_{1,RX}|^2  p_{x} \label{eq:p_si_f} \\[2mm]
	p_{SI,im} &= (E[|k_{BB}|^2 |k_{LNA}|^2 |\alpha_{0}|^2 |g_{2,TX}|^2 |g_{1,RX}|^2 \nonumber\\
	&\times|h_{ch} - a|^2] + E[|k_{BB}|^2 |k^{\ast}_{LNA}|^2 |\alpha^{\ast}_{0}|^2 |g^{\ast}_{1,TX}|^2 \nonumber\\
	&\times|g_{2,RX}|^2|h_{ch}^{\ast} - a^{\ast}|^2] + 2E[Re\{ |k_{BB}|^2 k^2_{LNA} \alpha^2_{0}\nonumber\\
	&\times g_{1,TX} g_{2,TX} g_{1,RX} g^{\ast}_{2,RX}(h_{ch} - a)^2 \}]) p_{x} \nonumber\\
	&= (|k_{BB}|^2 |k_{LNA}|^2 |\alpha_{0}|^2 |g_{2,TX}|^2 |g_{1,RX}|^2\nonumber\\
	&\times E[|h_{ch} - a|^2] + |k_{BB}|^2 |k^{\ast}_{LNA}|^2 |\alpha^{\ast}_{0}|^2|g^{\ast}_{1,TX}|^2\nonumber\\
	&\times |g_{2,RX}|^2 E[|h_{ch}^{\ast} - a^{\ast}|^2] + 2Re\{ |k_{BB}|^2 k^2_{LNA} \alpha^2_{0} \nonumber\\
	&\times g_{1,TX} g_{2,TX} g_{1,RX} g^{\ast}_{2,RX} E[(h_{ch} - a)^2] \}) p_{x} \nonumber\\
	&= (|k_{BB}|^2 |k_{LNA}|^2 |\alpha_{0}|^2 |a_{ant}|^2 |a_{RF}|^2\nonumber\\
	&\times (|g_{2,TX}|^2 |g_{1,RX}|^2 + |g_{1,TX}|^2 |g_{2,RX}|^2) p_{x} \label{eq:p_si_im_f} \\[2mm]
	p_{IMD} &= |k_{BB}|^2 |k_{LNA}|^2 |\alpha_{1}|^2 |g_{1,RX}|^2\nonumber\\
	&\times |a_{ant}|^2 |a_{RF}|^2 p_{x,IMD} \label{eq:p_imd_f} \\[2mm]
	p_{IMD,im} &= |k_{BB}|^2 |k_{LNA}|^2 |\alpha_{1}|^2 |g_{2,RX}|^2\nonumber\\
	&\times |a_{ant}|^2 |a_{RF}|^2 p_{x,IMD} \label{eq:p_imd_im_f}\\[2mm]
	p_{noise} &= F |k_{BB}|^2 |k_{LNA}|^2 |g_{1,RX}|^2 p_{th} \label{eq:p_n_f}\\[2mm]
	p_{noise,im} &= F |k_{BB}|^2 |k_{LNA}|^2 |g_{2,RX}|^2 p_{th} \label{eq:p_n_im_f}\\[2mm]
	p_{q} &= \frac{p_{AD}}{\mathit{snr}_{ADC}} \text{.} \label{eq:p_q_f}
\end{align}
In the above equations it is assumed that the complex gains of the different RF components are static and deterministic, whereas the error of the RF channel estimate is assumed to be a circular random variable, as explained earlier. Furthermore, the final term of the first equation for $p_{SI,im}$ can be omitted, as $E[(h_{ch} - a)^2] = 0$ due to the circularity assumption. The above set of derived formulas in \eqref{eq:p_si_f}--\eqref{eq:p_q_f} can now be used to evaluate the powers of different distortion terms at detector input, and in particular how they depend on the transmit power, antenna isolation, active RF cancellation and linear digital cancellation as well as on the transceiver RF imperfections.

\subsubsection*{System Calculations Example}

To illustrate the relative strengths of the different signal components, typical component parameters are chosen, and \eqref{eq:p_si_f}--\eqref{eq:p_q_f} are used to determine the corresponding power levels, which will then be shown for a specified transmit power range. The used parameters are listed in Table~\ref{table:system_parameters}, and they correspond to a typical wideband transceiver with low-cost mass-product components. The value for IRR, describing the image attenuation, is chosen based on 3GPP LTE specifications \cite{LTE_specs}. The baseband VGA of the receiver chain is assumed to match the total waveform dynamics at the ADC input to the available voltage range.

\begin{table}[!t]
\renewcommand{\arraystretch}{1.3}
\caption{Example system level parameters for the full-duplex transceiver.}
\label{table:system_parameters}
\centering
\begin{tabular}{|c||c|}
\hline
\textbf{Parameter} & Value\\
\hline
Bandwidth & 12.5 MHz\\
\hline
Thermal noise floor at receiver input & -103.0 dBm\\
\hline
Receiver noise figure & 4.1 dB\\
\hline
SNR requirement at detector input & 10 dB \\
\hline
Sensitivity level & -88.9 dBm\\
\hline
Power of the received signal & -83.9 dBm\\
\hline
Transmit power & varied\\
\hline
PA gain & 27 dB\\
\hline
PA IIP3 & 20 dBm\\
\hline
Antenna separation & 40 dB\\
\hline
RF cancellation & 30 dB\\
\hline
LNA gain & 25 dB \\
\hline
IQ mixer gain (RX and TX) & 6 dB\\
\hline
IRR (RX and TX) & 25 dB\\
\hline
VGA gain (RX) & 1--51 dB\\
\hline
ADC bits & 12\\
\hline
ADC P-P voltage range & 4.5 V\\
\hline
PAPR & 10 dB\\
\hline
\end{tabular}
\end{table}

In addition to the distortion powers, the power of the received signal of interest at detector input, denoted by $p_{SOI}$, is also shown in the figures as a reference, although it is not included in the signal model. This is done in order to be able to put the various distortion powers to proper context. The power of the signal of interest is chosen to be 5 dB above the sensitivity level at the input of the receiver chain. As the receiver sensitivity is defined for a 10 dB thermal noise SNR at detector input, this means that the signal of interest will be 15 dB above the thermal noise floor in the digital domain, which is then also the detector input SINR if no SI is present.

\begin{figure}[!t]
\centering
\includegraphics[width=\columnwidth]{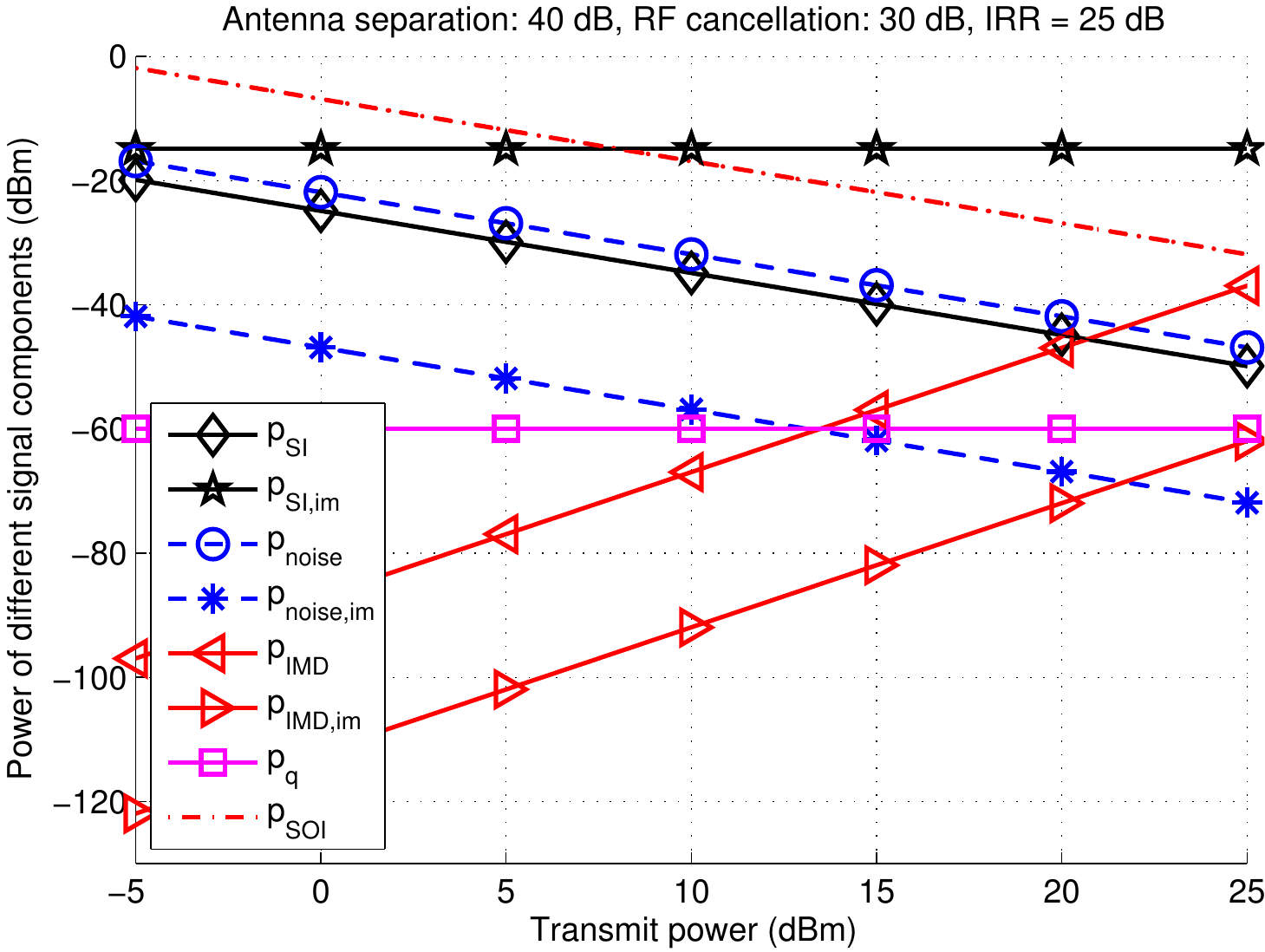}
\caption{The power levels of the different signal components after linear digital cancellation.}
\label{fig:syscalc}
\end{figure}

The power levels of the different signal components after linear digital cancellation, for transmit powers from -5 dBm to 25 dBm, are shown in Fig.~\ref{fig:syscalc}. The amount of digital cancellation is chosen so that linear SI is attenuated below the thermal noise floor. This has been observed to be close to the true performance of digital cancellation under realistic conditions \cite{Korpi13,Bharadia13}. In this example, this requires 27--57 dB of linear digital cancellation, depending on the transmit power.

From Fig.~\ref{fig:syscalc} it can be observed that the conjugate SI signal is clearly the most dominant distortion under a wide range of transmit powers. Actually, with transmit powers above 9 dBm, the power of the conjugate SI is even more powerful than the power of the signal of interest. Thus, with the chosen parameters, the conjugate SI is seriously degrading the achievable SINR of the full-duplex transceiver, which motivates the study of possible methods for attenuating it. Now, the ideal SINR of 15 dB is unreachable with the whole transmit power range from -5 dBm onwards, due to the powerful conjugate SI.

To investigate a different scenario, the amount of antenna separation is next assumed to be 30 dB, and the amount of RF cancellation 20 dB, which are achievable figures, even with very small antenna distance and low quality components for RF cancellation \cite{Choi10,Sahai11}. In addition, the IRR of the IQ mixers is increased to 35 dB to model the effect of using expensive, higher quality IQ mixers. Again, the amount of digital cancellation is chosen so that it attenuates the linear SI below the thermal noise, now requiring 47--77 dB of linear digital SI attenuation.

\begin{figure}[!t]
\centering
\includegraphics[width=\columnwidth]{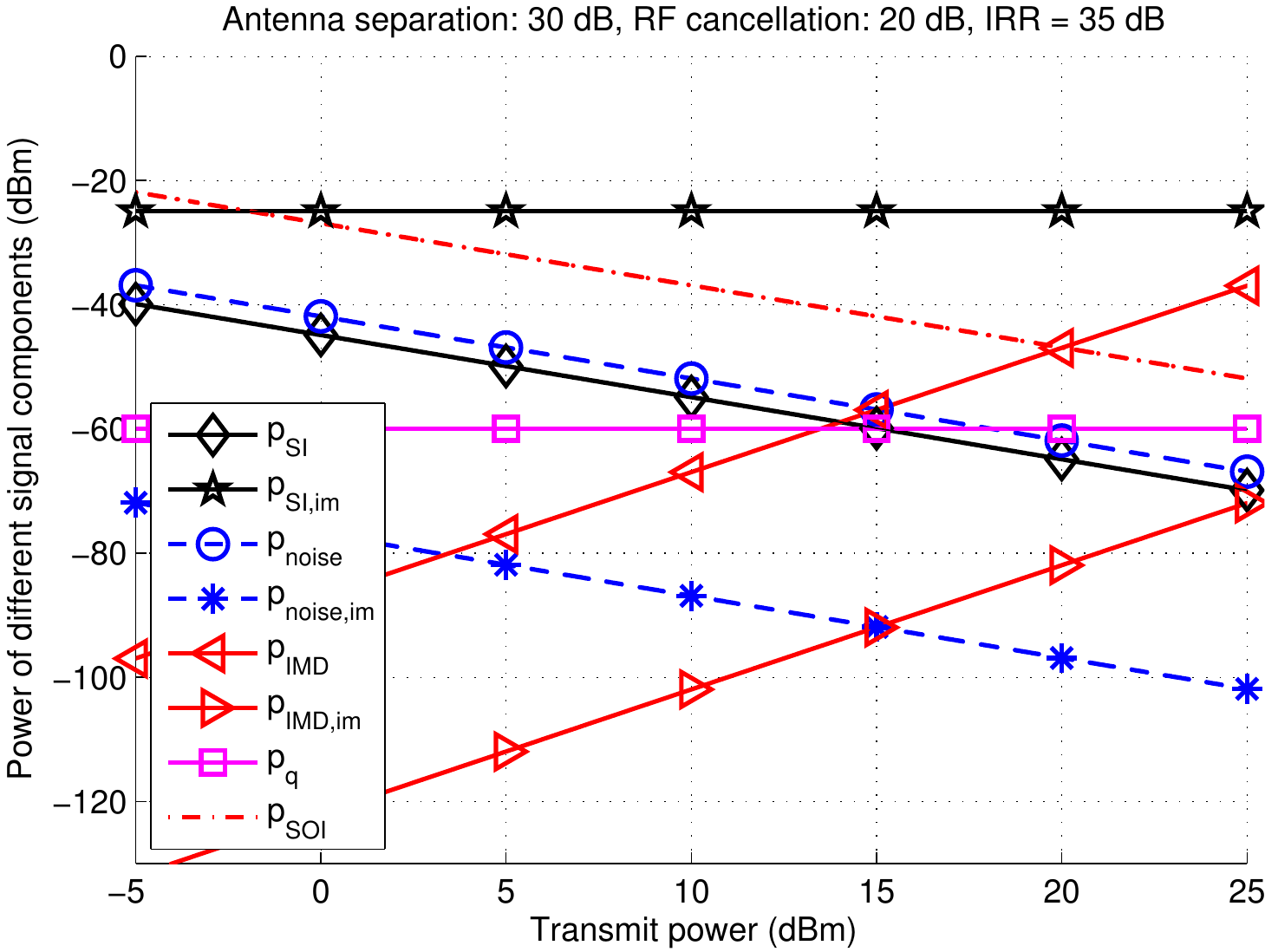}
\caption{The power levels of the different signal components after linear digital cancellation with the altered parameters.}
\label{fig:syscalc2}
\end{figure}

The resulting power levels are shown in Fig.~\ref{fig:syscalc2}. Now it can be observed that the power of the conjugate SI is even higher with respect to the power of the signal of interest, due to less analog SI cancellation. Thus, with more pessimistic values for antenna separation and RF cancellation, the effect of conjugate SI is very severe, even when using higher quality IQ mixers. It should be noted, however, that with transmit powers above 15 dBm, also the IMD produced by the PA can be observed to be a significant factor. This indicates that, in order to achieve higher transmit powers with these parameters, attenuating only the linear and conjugate SI may not be sufficient, as also the nonlinear distortion will degrade the signal quality.

\section{Proposed Widely-Linear Digital Cancellation}
\label{sec:wl_canc}

It was observed in the previous section that, with typical component parameters, conjugate SI is the dominating source of distortion after classical linear digital cancellation. Thus, the performance of the analyzed full-duplex transceiver can be enhanced by attenuating also the SI image component in the cancellation processing. This can be done by utilizing widely-linear digital SI cancellation, the principle of which is proposed and formulated in this section.

In order to focus on the IQ image induced problematics, we assume below that the IMD produced by the PA, alongside with the conjugate IMD, are clearly weaker than the conjugate SI. This is also clearly visible in Fig.~\ref{fig:syscalc}. We wish to acknowledge, though, that under certain circumstances, also the PA-induced nonlinear SI component must be further suppressed \cite{Anttila13,Ahmed13,Bharadia13}, but below we focus for simplicity on the dominating conjugate SI stemming from the IQ imbalances. A combined canceller for suppressing both the IMD and the conjugate SI is left for future work.

\subsection{Widely-Linear Cancellation Principle}

Our starting point is the fundamental signal model in \eqref{eq:y_c_compact}, which we can first re-write as
\begin{align*}
	y_{ADC}(n) &= h_{1}(n) \star x(n) + h_{2}(n) \star x^{\ast}(n) + z(n) \text{,}
\end{align*}
where
\begin{align*}
	h_{1}(n) &= k_{BB} k_{LNA} \alpha_{0} g_{1,TX}(n) \star g_{1,RX}(n) \\
	&\star (h_{ch}(n) - a(n)) +k_{BB} k^{\ast}_{LNA} \alpha^{\ast}_{0} g^{\ast}_{2,TX}(n)\\
	&\star g_{2,RX}(n) \star (h_{ch}^{\ast}(n) - a^{\ast}(n)) \text{,}\\[1.5mm]
	h_{2}(n) &= k_{BB} k_{LNA} \alpha_{0} g_{2,TX}(n) \star g_{1,RX}(n)\\
	&\star (h_{ch}(n) - a(n)) +k_{BB} k^{\ast}_{LNA} \alpha^{\ast}_{0} g^{\ast}_{1,TX}(n)\\
	&\star g_{2,RX}(n) \star (h_{ch}^{\ast}(n) - a^{\ast}(n)) \text{,}
\end{align*}
and $z(n)$ denotes in general the sum of all other signal terms, most notably thermal noise and quantization noise, and also PA-induced nonlinear distortion if present. Notice that here the general case of frequency-dependent antenna coupling and frequency-dependent IQ imbalances is considered, without any approximations.

The cancellation of the conjugate SI can be done in a similar manner as the cancellation of the direct component in \eqref{eq:y_c_dc}. In this case, however, the known transmitted signal must first be conjugated, and the conjugated samples are then filtered with the channel estimate of the effective image channel, $h_2(n)$, to produce the cancellation signal. If the estimate for the channel of the conjugate SI signal is denoted by $w_2(n)$, corresponding to the notation in Fig.~\ref{fig:baseband_model}, then the total signal after the cancellation stage can be expressed as
\begin{align}
	y_{WLDC}(n) &= y_{ADC}(n) - w_1(n) \star x(n) - w_2(n) \star x^{\ast}(n)\nonumber\\
	&= h_{1}(n) \star x(n) - w_1(n) \star x(n) + h_{2}(n) \star x^{\ast}(n)\nonumber\\
	&- w_2(n) \star x^{\ast}(n) + z(n) \nonumber\\
	&= (h_{1}(n) - w_1(n)) \star x(n)\nonumber\\
	&+ (h_{2}(n) - w_2(n))\star x^{\ast}(n) + z(n) \label{eq:y_c_all_canc}
\end{align}
Thus, by estimating $h_{2}(n)$, and convolving the conjugated transmit samples with it, a cancellation signal for the conjugate SI can be obtained. This type of cancellation procedure, which takes into account also the mirror image component, in addition to the direct linear component, is referred to as widely-linear (WL) digital cancellation as it is processing both the direct transmit data as well as its complex conjugate \cite{Schreier10,Picinbono95}.

\subsection{Widely-Linear Least-Squares Parameter Estimation}

The parameters required for the proposed WL digital cancellation can be obtained with ordinary least-squares estimation where the samples of the observed signal $y_{ADC}(n)$ serve as the reference and two estimation filters are fitted to it through the known transmit data $x(n)$ and its complex conjugate $x^{\ast}(n)$. In the following, to allow a more articulate mathematical description of the parameter estimation and the corresponding cancellation procedure, vector-matrix notations are used.

At digital baseband, the observed signal is $y_{ADC}(n) = h_1(n) \star x(n) + h_2(n) \star x^{\ast}(n) + z(n)$. Stacking the signals at hand into column vectors over an observation period of $N$ samples, this can be written as $\mathbf{y}_{ADC} = \mathbf{H}_1 \mathbf{x} + \mathbf{H}_2 \mathbf{x}^{\ast} + \mathbf{z}$, where $\mathbf{H}_1$ and $\mathbf{H}_2$ denote convolution matrices. Due to the commutability of the convolution operation, $\mathbf{y}_{ADC}$ can also be expressed as
\begin{align}
	\mathbf{y}_{ADC} &= \mathbf{X} \mathbf{h}_1 + \mathbf{X^{\ast}} \mathbf{h}_2 + \mathbf{z} = \begin{bmatrix} \mathbf{X} & \mathbf{X}^{\ast}\end{bmatrix}  \begin{bmatrix} 
	\mathbf{h}_1\\
	\mathbf{h}_2
	\end{bmatrix} + \mathbf{z}\nonumber\\ 
	&= \mathbf{X}_{aug} \mathbf{h}_{aug} + \mathbf{z} \label{eq:y_vec} \text{,}
\end{align}
where \mbox{\footnotesize$\mathbf{y}_{ADC} = [y_{ADC}(M-1) \text{ } y_{ADC}(M) \cdots y_{ADC}(N-K-1)]^{T}$}, and the covariance windowed convolution data matrix $\mathbf{X}$, with $K$ upper rows removed, is of the form
\begin{align*}
	\mathbf{X} =  \begin{bmatrix}
  x(M+K-1) & x(M+K-2) & \cdots & x(K) \\
  x(M+K) & x(M+K-1) & \cdots & x(K+1) \\
  \vdots  & \vdots  & \ddots & \vdots  \\
  x(N-1)& x(N-2) & \cdots & x(N-M)
 \end{bmatrix}
\end{align*}
The corresponding matrix, with each entry complex-conjugated, is denoted by $\mathbf{X}^{\ast}$. Here, $M < N$ is the length of the FIR filters $\mathbf{h}_1$ and $\mathbf{h}_2$ modeling the linear and conjugated channel responses. Furthermore, to allow the modeling of the additional memory effects due to delay errors in the RF cancellation signal, as well as possible time misalignment between $\mathbf{y}_{ADC}$ and $\mathbf{X}$, the system is made non-causal, or delayed, by the removal of $K < M$ upper rows from $\mathbf{X}$ and $\mathbf{X}^{\ast}$ \cite{Hammi08}. In essence, this means that the first $K$ taps of $\mathbf{h}_1$ and $\mathbf{h}_2$ represent a non-causal or pre-cursor part of the impulse responses. Based on this, the augmented convolution matrix $\mathbf{X}_{aug}$ can be written as
\begin{align*}
	&\mathbf{X}_{aug} = \begin{bmatrix}\mathbf{X} & \mathbf{X}^{\ast}\end{bmatrix} \nonumber\\
	&= \left[\begin{smallmatrix}
  x(M+K-1) & \cdots & x(K) & x^{\ast}(M+K-1) & \cdots & x^{\ast}(K) \\
  x(M+K) & \cdots & x(K+1) & x^{\ast}(M+K) & \cdots & x^{\ast}(K+1) \\
  \vdots & \ddots & \vdots & \vdots  & \ddots & \vdots  \\
  x(N-1) & \cdots & x(N-M) & x^{\ast}(N-1) & \cdots & x^{\ast}(N-M)
 \end{smallmatrix}\right]
\end{align*}
while the augmented channel $\mathbf{h}_{aug}$ contains $\mathbf{h}_1$ and $\mathbf{h}_2$ stacked as $\mathbf{h}_{aug} = \begin{bmatrix}\mathbf{h}^T_1 & \mathbf{h}^T_2\end{bmatrix}^{T}$.

Using the previous notation, the least-squares estimator for the augmented channel $\mathbf{h}_{aug}$ can then be calculated as
\begin{align}
	\mathbf{\hat{h}}_{aug} = (\mathbf{X}_{aug}^{H} \mathbf{X}_{aug})^{-1} \mathbf{X}_{aug}^{H} \mathbf{y}_{ADC} \text{,} \label{eq:h_est}
\end{align}
where $(\cdot)^{H}$ denotes the Hermitian transpose. In practice, alternative computationally efficient and numerically stable methods, such as singular value decomposition (SVD) based computations, can be used to evaluate the pseudo-inverse in \eqref{eq:h_est}. From the augmented channel estimate, due to the stacking property, the individual channel estimates for the direct and image channel can then be obtained directly as
\begin{align*}
	\mathbf{\hat{h}}_{1} &= \begin{bmatrix} \mathbf{\hat{h}}_{aug}(0) & \mathbf{\hat{h}}_{aug}(1) & \cdots & \mathbf{\hat{h}}_{aug}(M-1)\end{bmatrix}^{T}\\
	\mathbf{\hat{h}}_{2} &= \begin{bmatrix} \mathbf{\hat{h}}_{aug}(M) & \mathbf{\hat{h}}_{aug}(M+1) & \cdots & \mathbf{\hat{h}}_{aug}(2M-1)\end{bmatrix}^{T}
\end{align*}
Finally, the actual WL cancellation coefficients are directly assigned through $w_1(n) = \hat{h}_1(n)$ and $w_2(n) = \hat{h}_2(n)$, as is evident from \eqref{eq:y_c_all_canc}. With good estimates for the direct and conjugated channels, both the linear SI and conjugate SI will be efficiently attenuated, as will be illustrated in Section~\ref{sec:simulations}.

\section{Performance Simulations and Examples}
\label{sec:simulations}

The performance of the proposed WL digital SI cancellation method, including the WL least-squares parameter estimation, is next assessed and illustrated using full-scale waveform simulations of a complete full-duplex transceiver during simultaneous transmission and reception. The waveform simulator is implemented with Matlab, where each component is modeled explicitly and actual OFDM signals are used to ensure that also the signals themselves are realistic. The overall structure of the simulated transceiver is similar to the model presented in Fig.~\ref{fig:block_diagram}. The nonlinearity of the PA is realized by modeling it with a memory polynomial, whose coefficients are derived based on the intercept points. This means that, when the signal is amplified, it will also be distorted in a realistic manner. The effect of IQ imbalance is modeled by introducing some amplitude and phase errors to the I- and Q-branches. These errors are chosen so that they produce the desired value of image attenuation. Also the analog-to-digital conversion is modeled explicitly as a uniform quantization process, which means that the effect of quantization noise is incorporated into the simulations. The effect of thermal noise is realized by summing a normally distributed random signal to the overall signal at the receiver input, after which its power is increased according to the component gains and noise figures. Note that, due to the realistic and detailed modeling of the transceiver chain, no simplifications are made in the simulations regarding the different nonidealities.

Essentially the same parameters are used in the simulations as in the system calculations example, presented in Table~\ref{table:system_parameters}. The additional parameters, describing the utilized OFDM waveform, are presented in Table~\ref{table:simul_param}. These parameters are in essence similar to WLAN specifications, and both the transmitted and received signals are generated according to them. Similar to the system calculations example, due to the chosen power level for the received signal, SNR at the input of the detector is 15 dB when there are no non-idealities in addition to thermal noise. This serves as the reference value for the measured SINR in assessing how well the total SI can be suppressed in the receiver chain. Namely, if the SI is perfectly cancelled, a SINR of 15 dB is achieved.

\begin{table}[!t]
\renewcommand{\arraystretch}{1.3}
\caption{Additional parameters for the full-duplex waveform simulator. Baseline parameters are listed in Table I.}
\label{table:simul_param}
\centering
\begin{tabular}{|c||c|}
\hline
\textbf{Parameter} & \textbf{Value}\\
\hline
Constellation & 16-QAM\\
\hline
Number of subcarriers & 64\\
\hline
Number of data subcarriers & 48\\
\hline
Guard interval & 25 \% of symbol length\\
\hline
Sample length & 15.625 ns\\
\hline
Symbol length & 4 $\mu$s\\
\hline
Oversampling factor & 4\\
\hline
\end{tabular}
\end{table}

In an individual realization, the SI coupling channel between the antennas is assumed to be static, and it is modelled as a line-of-sight component and two weak multipath components, delayed by one and two sample intervals. The average ratio between the power of the main component and the total power of the multipath components is chosen to be 35.8 dB, which is a realistic value for a SI coupling channel when the full-duplex transceiver is located indoors \cite{Duarte12}.

In the simulations, RF cancellation attenuates only the direct component, as assumed also, e.g., in \cite{Choi10,Jain11,Radunovic09}, whereas the attenuation of the weaker multipath components is done by the proposed digital cancellation algorithm after the ADC. In addition, some delay and amplitude errors are included in the RF cancellation signal to achieve the desired amount of SI attenuation, and to model the cancellation process in a realistic manner. The delay error is implemented as a fractional delay for the cancellation signal, whose value is set to roughly 10~\% of sample duration. Together with a small amplitude error, the amount of achieved RF cancellation can then be set to the desired value.

The proposed WL cancellation scheme is analyzed based on two metrics: the SINR at detector input, and the amount of achieved digital SI attenuation. In the simulations, the actual SI channel estimation is done during a specified training period, where there is no actual received signal of interest present. This corresponds to the analysis presented earlier in this paper. The hereby obtained cancellation filter coefficients are then applied in the actual receiver operation mode, with useful received signal present, where the proposed WL canceller output signal, with estimated coefficients, is subtracted from the total received signal. This naturally allows then also the measurement of the SINR at detector input.

Unlike in Section~\ref{sec:syscalc}, the amount of achieved digital SI attenuation is now redefined as the decrease in the power of both the linear SI and conjugate SI. It can thus be written as
{\small
\begin{align}
	&|a_{WLDC}|^2 = \frac{E[|y_{WLDC,SI}(n)|^2]}{E[|y_{ADC,SI}(n)|^2]} \nonumber\\
	&= \frac{E[|(h_1(n) - w_1(n)) \star x(n) + (h_2(n) - w_2(n)) \star x^{\ast}(n)|^2]}{E[|h_1(n) \star x(n) + h_2(n) \star x^{\ast}(n)|^2]} \label{eq:a_wldc}
\end{align}}%
This definition for digital SI attenuation is used in order to be able to compare the performances of classical linear and widely-linear models for digital cancellation. In the simulations, the amount of digital SI attenuation is then measured according to \eqref{eq:a_wldc}, using the coefficients that were estimated under regular noisy conditions. In these performance measurements, no simplifications are done in order to make sure that the obtained cancellation figures are as accurate as possible.

In the first simulations, transmit power is varied with 2 dB intervals, and the simulation is repeated for 1000 independent realizations for each transmit power. The average value of these runs for the SINR and the amount of digital cancellation is then used in the figures. The number of training samples used for cancellation parameter estimation ($N$) is fixed to 5000, and the value of $K$ is set to $1$, as the fractional delay of the RF cancellation signal may produce additional non-causal memory effects to the resulting signal \cite{Hammi08}. The length of the channel estimate ($M$) is set to 5, as it will allow the modeling of the multipath components and some of the additional memory effects of the SI channel.

\begin{figure}[!t]
\centering
\includegraphics[width=\columnwidth]{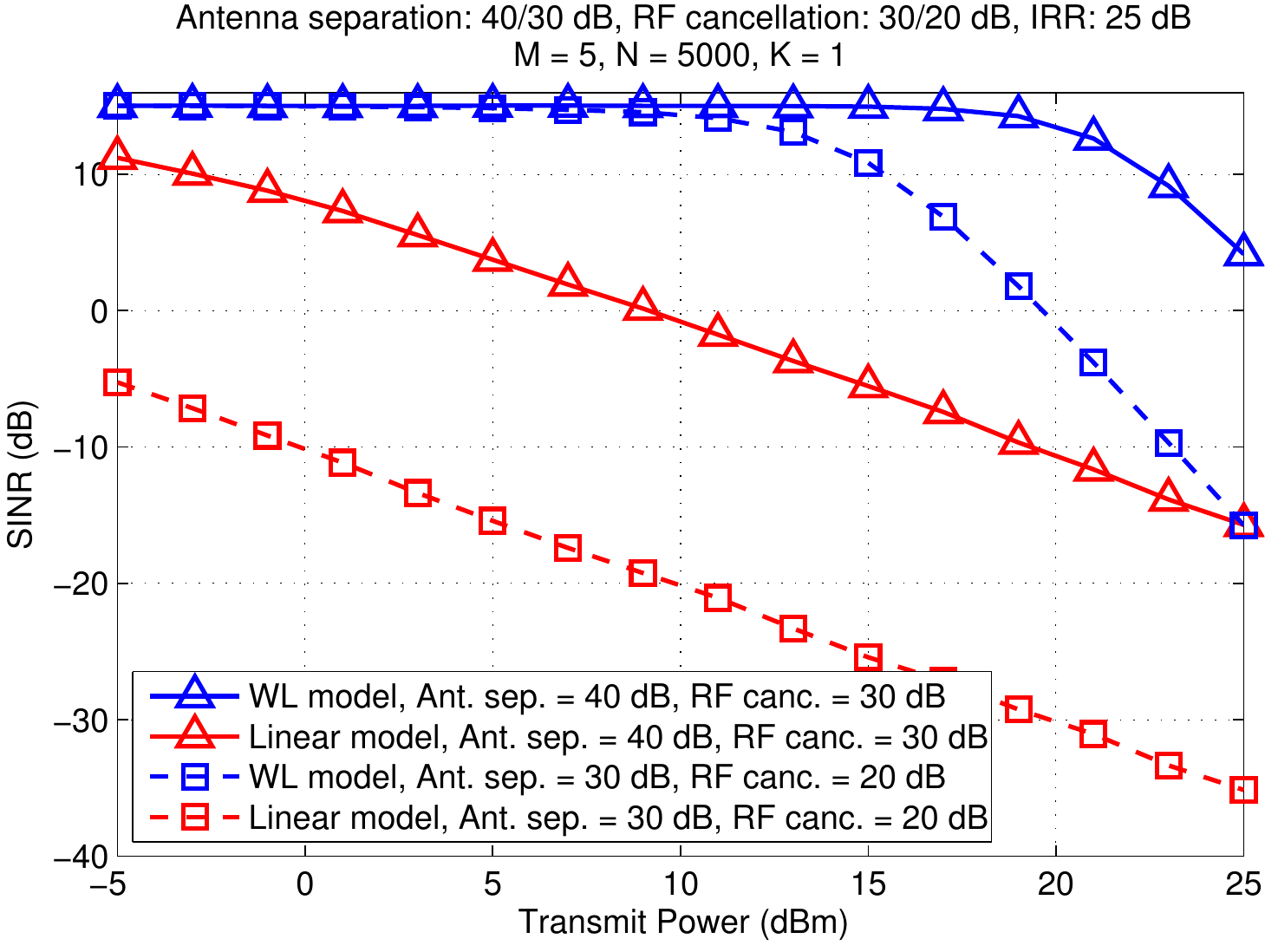}
\caption{The SINRs achieved with the proposed WL digital cancellation and with traditional linear digital cancellation. The figures are shown for higher and lower amounts of analog SI attenuation.}
\label{fig:sinr_ptx}
\end{figure}

In Fig.~\ref{fig:sinr_ptx}, the SINR with respect to transmit power is shown for both the proposed WL digital canceller, and for the traditional linear digital cancellation method. The latter corresponds to a situation, where the channel estimate is calculated with linear least squares as $\mathbf{\hat{h}}_1 = (\mathbf{X}^{H} \mathbf{X})^{-1} \mathbf{X}^{H} \mathbf{y}_{ADC}$, and $\mathbf{\hat{h}}_2 = \mathbf{0}$. The SINR curves are shown for two cases: in the first case, the amounts of antenna separation and RF cancellation are 30 dB and 40 dB, whereas in the second case they are only 30 dB and 20 dB, respectively.

When investigating the curves corresponding to the higher amount of analog SI attenuation (40 dB of antenna separation and 30 dB of RF cancellation) in Fig.~\ref{fig:sinr_ptx}, it can be observed that the SINR declines steadily when using only the classical linear model and linear SI cancellation. The effect of the conjugate SI increases with higher transmit powers, as its power is directly related to the power of the SI signal before digital cancellation. On the other hand, when using the proposed WL model and WL processing, it can be observed that the SINR remains essentially at the ideal level of 15 dB with transmit powers below 15 dBm. After this point, it is not possible to achieve the ideal SINR even with the WL digital canceller, as the SINR is decreased by the PA-induced IMD, whose power is increasing rapidly with higher transmit powers. Another factor contributing to the decrease in the SINR is the decreasing resolution of the signal of interest, caused by quantization noise. This is because ideal automatic gain control is assumed in the receiver, always matching the total ADC input waveform to the available ADC voltage range. Nevertheless, a significant improvement in the SINR can be achieved when the proposed WL digital cancellation method is used and also conjugate SI is attenuated in the digital domain.

\begin{figure}[!t]
\centering
\includegraphics[width=\columnwidth]{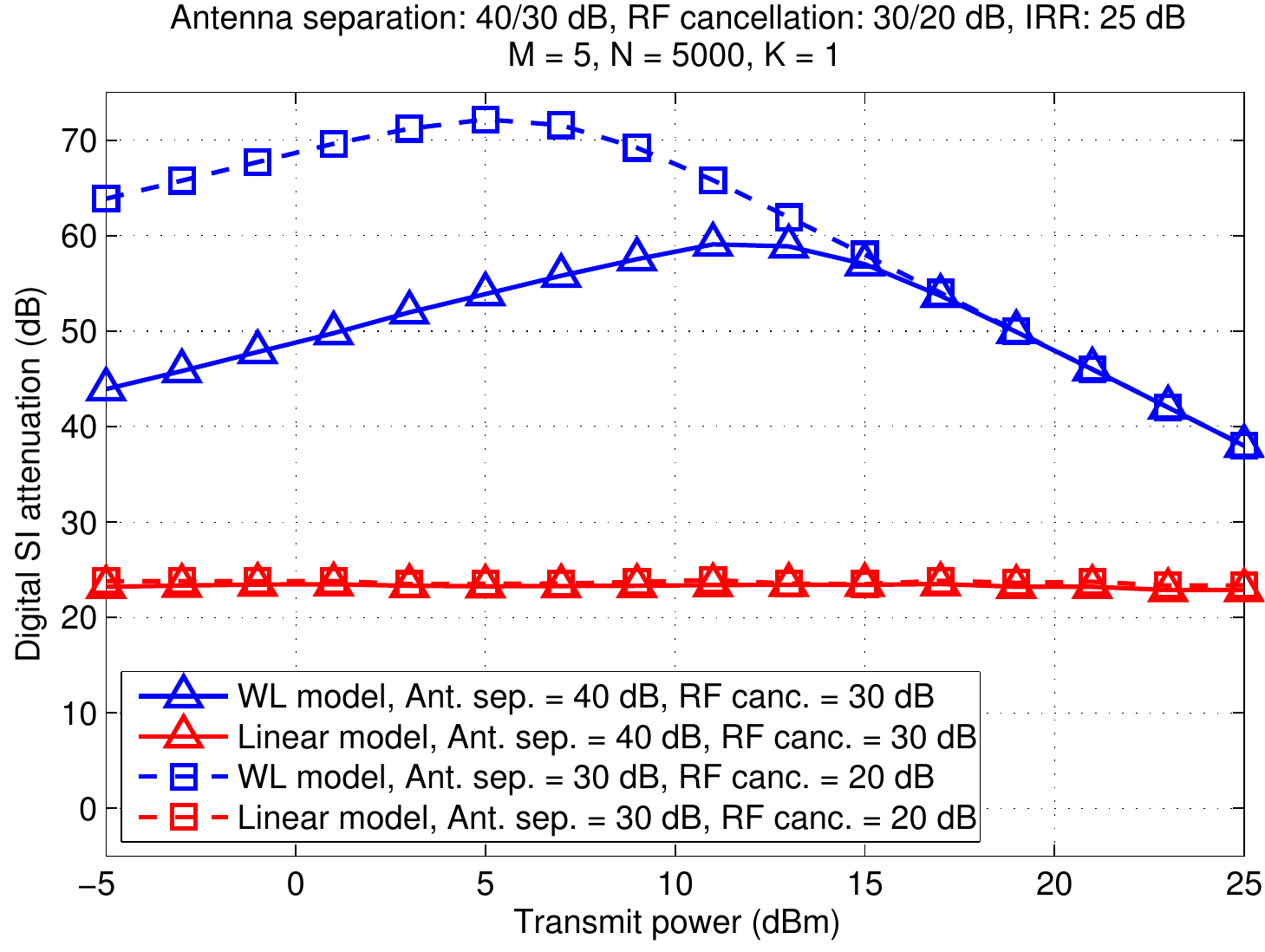}
\caption{The amounts of achieved digital cancellation using the proposed WL processing and using traditional linear processing. The figures are shown for higher and lower amounts of analog SI attenuation.}
\label{fig:digcanc_ptx}
\end{figure}

Since it might be in some instances quite optimistic to assume 40 dB of antenna separation and 30 dB of RF cancellation, in Fig.~\ref{fig:sinr_ptx} the SINRs are also shown for a situation where both antenna separation and RF cancellation are 10 dB lower. This will mean that the power of the total SI signal with respect to the signal of interest is significantly higher in the digital domain. Now it can be observed that, when using WL digital cancellation, the residual SI starts to degrade the SINR already with transmit powers above 7 dBm. This is caused by PA-induced IMD, which cannot be modeled by the proposed WL canceller, as discussed earlier. In addition, at this point quantization noise is also slightly decreasing the SINR. However, again, the improvement over traditional linear canceller is significant, as the SINR for the linear model is -5 dB already with the lowest considered transmit power, when assuming this amount of analog SI attenuation. Thus, the significance of using the proposed WL model for digital cancellation is further emphasized when having less analog SI attenuation. These observations also motivate towards joint cancellation of linear SI, conjugate SI as well as nonlinear SI due to PA. This forms the topic of our future work.

Figure~\ref{fig:digcanc_ptx} shows the amount of achieved digital SI attenuation for WL processing and traditional linear processing, corresponding to the SINR curves shown in Fig.~\ref{fig:sinr_ptx}. It can be observed that the amount of achieved digital SI attenuation with the linear model is below 25 dB, regardless of the amount of analog SI attenuation. On the other hand, the achieved SI attenuation with the WL model is as much as 35~dB or 50~dB higher, depending on the amount of antenna separation and RF cancellation. This is explained by the fact that, when using the linear model, the power of the conjugate SI is in essence the noise floor in the digital domain, and it determines the lowest achievable total SI power. Thus, the amount of achieved digital SI attenuation is limited by the power of the conjugate SI. For the WL model, however, this is not the case, and significantly higher values of SI cancellation can be achieved, limited only by the accuracy of the channel estimate.

It can also be observed from Fig.~\ref{fig:digcanc_ptx} that when the amount of analog SI attenuation is higher, the amount of achieved digital SI attenuation is lower, assuming that the transmit power is below 15 dBm. This is explained by the SNR of the total SI signal in the two scenarios: when the amount of analog SI attenuation is low, the power of the total SI signal with respect to the noise is higher in the digital domain, resulting in a more accurate channel estimate. This obviously means that more digital SI attenuation is achieved when the amount of analog SI attenuation is lower, which has also been observed in actual RF measurements \cite{Duarte10}.

\begin{figure}[!t]
\centering
\includegraphics[width=\columnwidth]{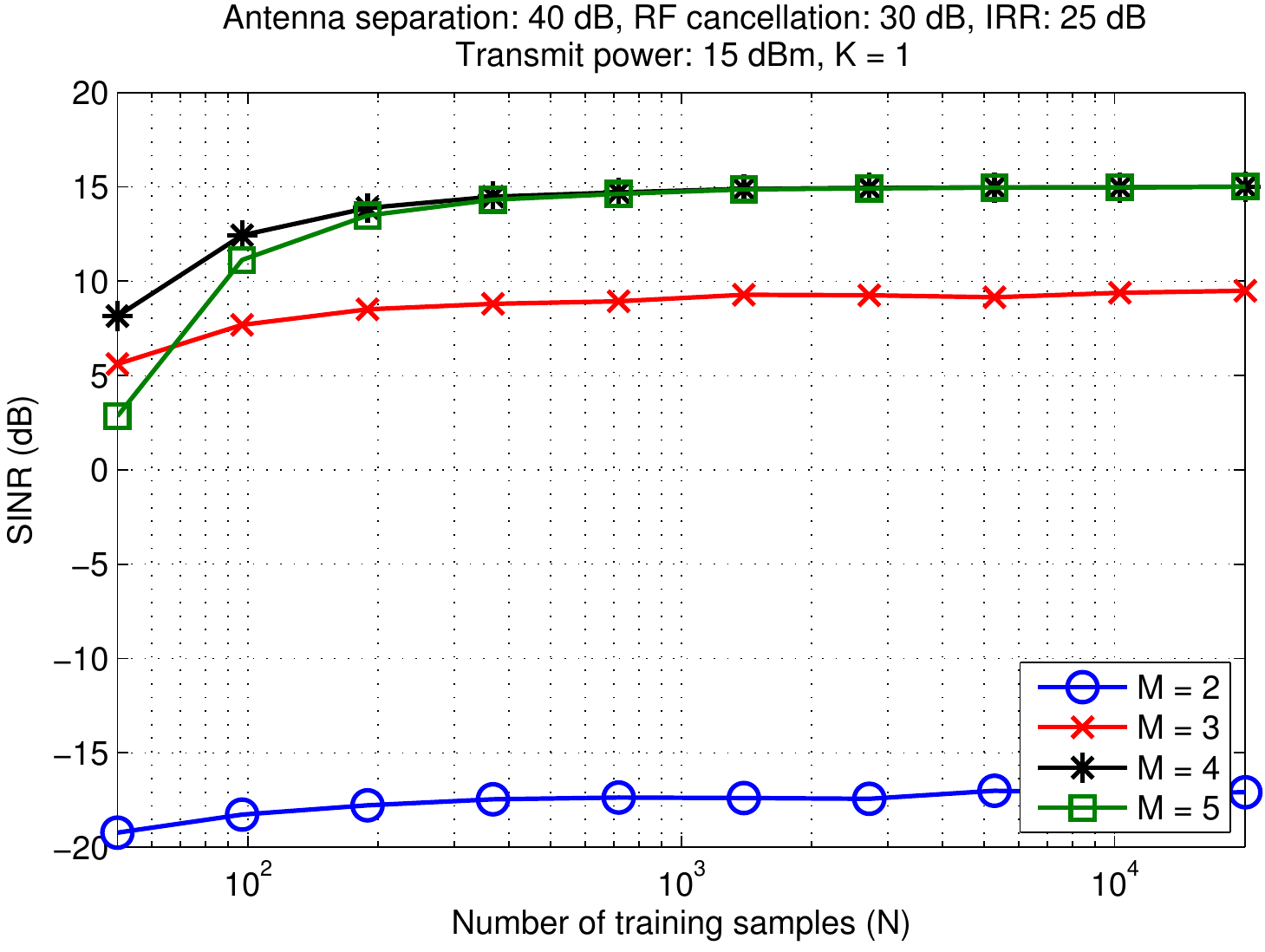}
\caption{The SINRs achieved with different lengths for the channel estimate filters, with respect to the number of training samples, using WL digital cancellation.}
\label{fig:sinr_mn}
\end{figure}

However, with transmit powers above 15 dBm, the amount of achievable digital SI attenuation is also limited by the IMD, as the nonlinear PA is hindering the parameter estimation accuracy at higher transmit powers. More specifically, the nonlinear distortion due to PA causes a bias to the estimates, as shown in the Appendix, and decreases their accuracy. Thus, with higher transmit powers, the performance of the WL digital cancellation decreases due to increasing bias in the estimation, as well as due to higher residual SI power caused by the IMD. Furthermore, according to \eqref{eq:h_1}--\eqref{eq:h_x_imd_conj}, IMD is attenuated in the analog domain by the same amount as the linear SI and conjugate SI signals, and hence its relative power level in the digital domain does not depend on the amount of analog SI attenuation. Thus, with higher transmit powers, when IMD is the limiting factor, same amount of digital SI attenuation is achieved with both sets of values for antenna separation and RF cancellation, as is evidenced by Fig.~\ref{fig:digcanc_ptx}.

Next, the used transmit power is fixed to 15 dBm, and WL model is employed in digital cancellation. The simulation is then run with different lengths of channel estimate filters, and with varying number of training samples, the amount of analog SI attenuation being the same in each case. The number of training samples is varied from 50 to 20000 with 10 logarithmically spaced points in between, and channel estimate filter lengths from 2 to 5 are considered. The value of $K$ is again set to 1 in all cases. The simulation is then repeated for 400 independent runs at each of these points, and the values are averaged.

\begin{figure}[!t]
\centering
\includegraphics[width=\columnwidth]{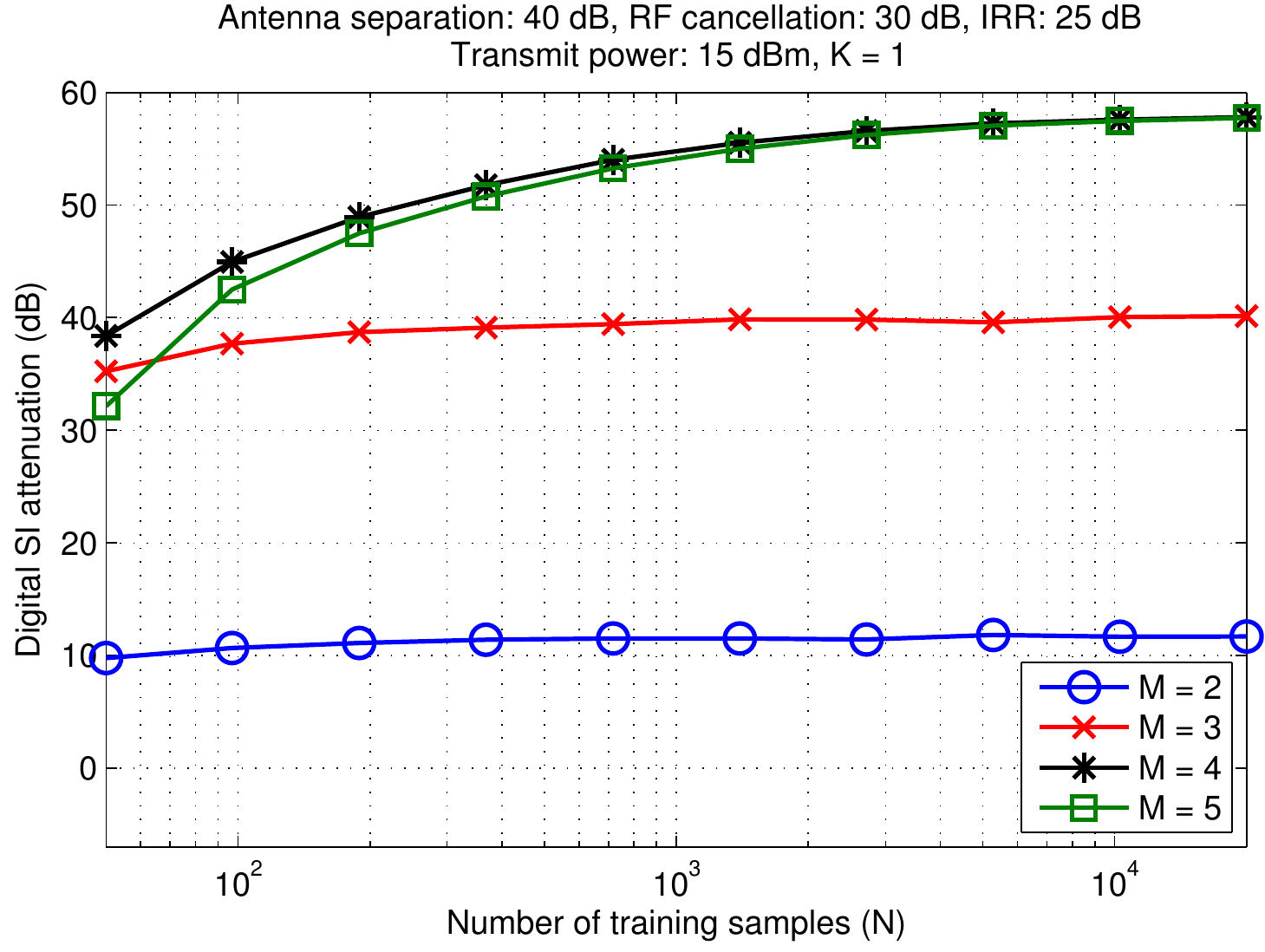}
\caption{The amount of digital SI attenuation achieved with different lengths for the channel estimate filters, with respect to the number of training samples, using WL digital cancellation.}
\label{fig:digcanc_mn}
\end{figure}

The SINRs obtained from these simulations are shown in Fig.~\ref{fig:sinr_mn}. It can be observed that if the value for $M$ is higher than or equal to $4$, there are no significant differences in the achieved SINRs, assuming that the number of training samples is at least 1000. The likely reason for this is that when $M \geq 4$, all the multipath components can be modelled by the channel estimate. This obviously suggests that the channel conditions must be carefully analyzed when choosing a suitable length for the channel estimate filter. Furthermore, the delay error of the RF cancellation signal does not seem to produce any significant causal taps to the total SI channel, as there are no discernible differences in the maximum achievable SINRs between $M=4$ and $M=5$.

Figure~\ref{fig:sinr_mn} also indicates that increasing the number of training samples is not likely to bring any benefits after a certain point, as the power of the total SI signal, consisting of both the linear and conjugate SI, is then attenuated well below the noise floor. With the chosen parameters, the value of the SINR saturates with approximately 3000 training samples, regardless of the length of the channel estimate filter. However, it can be observed that the length of the filter affects the minimum number of training samples required to achieve the saturation value of the SINR; with smaller values of $M$, less samples are required to achieve the highest possible SINR. The reason for this is the higher variance caused by estimating a larger number of coefficients using the same number of training samples.

The amount of achieved digital SI attenuation, corresponding to the SINRs of Fig.~\ref{fig:sinr_mn}, is shown in Fig.~\ref{fig:digcanc_mn}. The form of the curves is similar to the achieved SINRs, i.e., the achieved cancellation is lower with less training samples. Also, the maximum amount of digital SI attenuation is achieved when $M \geq 4$, indicating that this channel estimate length is sufficient to model the actual RF propagation channel.  However, this result is heavily dependent on the characteristics of the coupling channel between the antennas, and the accuracy of the delay matching of the RF cancellation signal. Under certain circumstances, longer filters might be needed to model the total SI channel with sufficient accuracy.

In addition, with this transmit power, the amount of digital SI attenuation can be observed to saturate to a value of approximately 58 dB, assuming that a sufficiently high number of training samples is used ($N > 10000$). This saturated value is mainly set by the IMD produced by the PA, which is caused by the fact that the PA distorts the SI waveform in a nonlinear manner, which cannot be modeled by the WL least squares estimation. This, on the other hand, will cause errors in the coefficient estimates, resulting in a lower amount of achieved digital cancellation, as was already discussed earlier. Nevertheless, as was observed from Fig.~\ref{fig:sinr_mn}, the highest possible SINR is achieved already with 1000--3000 training samples, depending on the value of $M$. Thus, based on Fig.~\ref{fig:digcanc_mn}, it can be concluded that, with transmit powers up to 15 dBm, approximately 55 dB of digital SI attenuation is sufficient to suppress the total SI signal well below the noise floor. Furthermore, our findings indicate that WL processing is required to be able to achieve digital SI attenuation of this magnitude.

\section{Conclusion}
\label{sec:conc}

In this paper, we have proposed a novel method for compensating the image component of the self-interference (SI) signal, caused by IQ imbalances in the transmitter and receiver IQ mixers. To the best of our knowledge, this is the first time this problem has been addressed in the literature at self-interference waveform and digital cancellation levels. With fundamental system calculations, it was first shown that the image component of the SI signal will affect the performance of a full-duplex transceiver by significantly decreasing the maximum achievable SINR. Then, by using the developed widely-linear least-squares parameter estimation and the proposed widely-linear digital cancellation, we showed that this image component can also be attenuated in the digital domain, and thus the decrease in the SINR can be prevented. These claims were affirmed with extensive full waveform simulations, which demonstrate that the proposed method will significantly improve the performance of a typical full-duplex transceiver. Future work will focus on developing joint augmented nonlinear digital cancellation processing that is able to suppress the classical linear SI, nonlinear SI produced by the transmitter power amplifier, and their image components due to IQ imaging.

\appendix[Analysis of Bias]

Direct substitution of \eqref{eq:y_vec} into \eqref{eq:h_est} yields $\mathbf{\hat{h}}_{aug} = \mathbf{h}_{aug} + (\mathbf{X}^H_{aug} \mathbf{X}_{aug})^{-1} \mathbf{X}^H_{aug} \mathbf{z} = \mathbf{h}_{aug} + \mathbf{e}$. To analyze the estimation error $\mathbf{e}$ analytically, we resort to the simplified frequency-independent model presented in Subsection~\ref{sec:syscalc}, that is $y_{ADC}(n) \approx h_1 x(n) + h_2 x^{\ast}(n) + h_{IMD} x_{IMD}(n) + u(n)$, where $x_{IMD}(n) = x^{TX}_{IQ}(n) |x^{TX}_{IQ}(n)|^2$ and $x_{IQ}^{TX}(n)$ is as defined in \eqref{eq:x_tx}, implying that $\mathbf{z} = h_{IMD} \mathbf{x}_{IMD} + \mathbf{u}$. Furthermore, with relatively large sample size $N$, $\mathbf{X}^H_{aug} \mathbf{X}_{aug}$ can be approximated with ensemble augmented covariance matrix $\mathbf{R}_{x,aug}$ of transmit data $x(n)$ \cite{Schreier10}, that is
\begin{align*}
\mathbf{R}_{x,aug} = \begin{bmatrix}
  r & c^{\ast}\\
  c & r
 \end{bmatrix} = \begin{bmatrix}
  r & 0\\
  0 & r
 \end{bmatrix} \approx \frac{1}{N} \mathbf{X}^H_{aug} \mathbf{X}_{aug} \text{,}
\end{align*}
where $r = E[|x(n)|^2]$ and $c = E[x(n)^2] = 0$ assuming $x(n)$ is 2nd-order circular. Hence, the estimation error can be written as
\begin{align*}
	\mathbf{e} \approx h_{IMD} N^{-1} \mathbf{R}^{-1}_{x,aug} \mathbf{X}^H_{aug} \mathbf{x}_{IMD} + N^{-1} \mathbf{R}^{-1}_{x,aug} \mathbf{X}^H_{aug} \mathbf{u}
\end{align*}
Due to the dependence of $\mathbf{X}_{aug}$ and $\mathbf{x}_{IMD}$, the average estimation error is nonzero, i.e., the estimator is biased. Assuming that $x(n)$ is also 4th-order circular (i.e., $E[x(n)^4] = E[x(n)^3 x^{\ast}(n)] = 0 $), this can be shown directly as
\begin{align*}
&E[h_{IMD} N^{-1} \mathbf{R}^{-1}_{x,aug} \mathbf{X}^H_{aug} \mathbf{x}_{IMD}]\nonumber\\
&= h_{IMD} N^{-1} \mathbf{R}^{-1}_{x,aug} E[\mathbf{X}^H_{aug} \mathbf{x}_{IMD}]\nonumber\\
&\approx h_{IMD} N^{-1} \mathbf{R}^{-1}_{x,aug} \begin{bmatrix}
  g_{1,TX} |g_{1,TX}|^2 N E[|x(n)|^4]\\
  2 g_{2,TX} |g_{1,TX}|^2 N E[|x(n)|^4]
 \end{bmatrix} \nonumber\\
  &= \frac{h_{IMD} |g_{1,TX}|^2 E[|x(n)|^4]}{r}\begin{bmatrix}
  g_{1,TX}\\
  2 g_{2,TX}
 \end{bmatrix} \nonumber\\
 &\neq \mathbf{0} \text{ q.e.d.} \nonumber
\end{align*}
Thus, as the above derivation shows, the IMD of the nonlinear transmitter power amplifier causes a bias to the estimator.

\bibliographystyle{./IEEEtran}
\bibliography{./IEEEabrv,./IEEEref}

\begin{thebibliography}{10}
\providecommand{\url}[1]{#1}
\csname url@samestyle\endcsname
\providecommand{\newblock}{\relax}
\providecommand{\bibinfo}[2]{#2}
\providecommand{\BIBentrySTDinterwordspacing}{\spaceskip=0pt\relax}
\providecommand{\BIBentryALTinterwordstretchfactor}{4}
\providecommand{\BIBentryALTinterwordspacing}{\spaceskip=\fontdimen2\font plus
\BIBentryALTinterwordstretchfactor\fontdimen3\font minus
  \fontdimen4\font\relax}
\providecommand{\BIBforeignlanguage}[2]{{%
\expandafter\ifx\csname l@#1\endcsname\relax
\typeout{** WARNING: IEEEtran.bst: No hyphenation pattern has been}%
\typeout{** loaded for the language `#1'. Using the pattern for}%
\typeout{** the default language instead.}%
\else
\language=\csname l@#1\endcsname
\fi
#2}}
\providecommand{\BIBdecl}{\relax}
\BIBdecl

\bibitem{Radunovic09}
B.~Radunovic, D.~Gunawardena, P.~Key, A.~Proutiere, N.~Singh, V.~Balan, and
  G.~DeJean, ``Rethinking indoor wireless mesh design: Low power, low
  frequency, full-duplex,'' in \emph{Proc. Fifth IEEE Workshop on Wireless Mesh
  Networks}, Jun. 2010, pp. 1--6.

\bibitem{Choi10}
J.~I. Choi, M.~Jain, K.~Srinivasan, P.~Levis, and S.~Katti, ``Achieving single
  channel full duplex wireless communication,'' in \emph{Proc. 16th Annual
  International Conference on Mobile Computing and Networking}, Sep. 2010, pp.
  1--12.

\bibitem{Jain11}
M.~Jain, J.~I. Choi, T.~Kim, D.~Bharadia, S.~Seth, K.~Srinivasan, P.~Levis,
  S.~Katti, and P.~Sinha, ``Practical, real-time, full duplex wireless,'' in
  \emph{Proc. 17th Annual International Conference on Mobile computing and
  Networking}, Sep. 2011, pp. 301--312.

\bibitem{Sahai11}
A.~Sahai, G.~Patel, and A.~Sabharwal, ``Pushing the limits of full-duplex:
  Design and real-time implementation,'' Department of Electrical and Computer
  Engineering, Rice University, Technical Report TREE1104, Jul. 2011.

\bibitem{Duarte10}
M.~Duarte and A.~Sabharwal, ``Full-duplex wireless communications using
  off-the-shelf radios: Feasibility and first results,'' in \emph{Proc. 44th
  Asilomar Conference on Signals, Systems, and Computers}, Nov. 2010, pp.
  1558--1562.

\bibitem{Day12}
B.~Day, A.~Margetts, D.~Bliss, and P.~Schniter, ``Full-duplex bidirectional
  {MIMO}: Achievable rates under limited dynamic range,'' \emph{IEEE
  Transactions on Signal Processing}, vol.~60, no.~7, pp. 3702--3713, Jul.
  2012.

\bibitem{Duarte12}
M.~Duarte, C.~Dick, and A.~Sabharwal, ``Experiment-driven characterization of
  full-duplex wireless systems,'' \emph{IEEE Transactions on Wireless
  Communications}, vol.~11, no.~12, pp. 4296--4307, Dec. 2012.

\bibitem{Duarte122}
M.~Duarte, A.~Sabharwal, V.~Aggarwal, R.~Jana, K.~Ramakrishnan, C.~Rice, and
  N.~Shankaranarayanan, ``Design and characterization of a full-duplex
  multiantenna system for {WiFi} networks,'' \emph{IEEE Transactions on
  Vehicular Technology}, vol.~63, no.~3, pp. 1160--1177, Mar. 2014.

\bibitem{Knox12}
M.~E. Knox, ``Single antenna full duplex communications using a common
  carrier,'' in \emph{Proc. IEEE 13th Annual Wireless and Microwave Technology
  Conference}, Apr. 2012, pp. 1--6.

\bibitem{Cox13}
C.~Cox and E.~Ackerman, ``Demonstration of a single-aperture, full-duplex
  communication system,'' in \emph{Proc. Radio and Wireless Symposium}, Jan.
  2013, pp. 148--150.

\bibitem{Phungamngern13}
N.~Phungamngern, P.~Uthansakul, and M.~Uthansakul, ``Digital and {RF}
  interference cancellation for single-channel full-duplex transceiver using a
  single antenna,'' in \emph{Proc. 10th International Conference on Electrical
  Engineering/Electronics, Computer, Telecommunications and Information
  Technology (ECTI-CON)}, May 2013, pp. 1--5.

\bibitem{Everett13}
E.~Everett, A.~Sahai, and A.~Sabharwal, ``Passive self-interference suppression
  for full-duplex infrastructure nodes,'' \emph{IEEE Transactions on Wireless
  Communications}, vol.~13, no.~2, pp. 680--694, Feb. 2014.

\bibitem{Aryafar12}
E.~Aryafar, M.~A. Khojastepour, K.~Sundaresan, S.~Rangarajan, and M.~Chiang,
  ``{MIDU}: enabling {MIMO} full duplex,'' in \emph{Proc. 18th Annual
  International Conference on Mobile Computing and Networking}, Aug. 2012, pp.
  257--268.

\bibitem{Bharadia13}
D.~Bharadia, E.~McMilin, and S.~Katti, ``Full duplex radios,'' in
  \emph{SIGCOMM'13}, Aug. 2013.

\bibitem{McMichael12}
J.~McMichael and K.~Kolodziej, ``Optimal tuning of analog self-interference
  cancellers for full-duplex wireless communication,'' in \emph{Proc. 50th
  Annual Allerton Conference on Communication, Control, and Computing}, Oct.
  2012, pp. 246--251.

\bibitem{Lee13}
J.-H. Lee, ``Self-interference cancelation using phase rotation in full-duplex
  wireless,'' \emph{IEEE Transactions on Vehicular Technology}, vol.~62, no.~9,
  pp. 4421--4429, Nov. 2013.

\bibitem{Anttila13}
L.~Anttila, D.~Korpi, V.~Syrj\"{a}l\"{a}, and M.~Valkama, ``Cancellation of
  power amplifier induced nonlinear self-interference in full-duplex
  transceivers,'' in \emph{Proc. 47th Asilomar Conference on Signals, Systems
  and Computers}, Nov. 2013, pp. 1193--1198.

\bibitem{Ahmed13}
E.~Ahmed, A.~M. Eltawil, and A.~Sabharwal, ``Self-interference cancellation
  with nonlinear distortion suppression for full-duplex systems,'' in
  \emph{Proc. 47th Asilomar Conference on Signals, Systems and Computers}, Nov.
  2013, pp. 1199--1203.

\bibitem{Riihonen13}
T.~Riihonen, M.~Vehkapera, and R.~Wichman, ``Large-system analysis of rate
  regions in bidirectional full-duplex {MIMO} link: Suppression versus
  cancellation,'' in \emph{Proc. 47th Annual Conference on Information Sciences
  and Systems (CISS)}, Mar. 2013, pp. 1--6.

\bibitem{Riihonen1222}
T.~Riihonen, P.~Mathecken, and R.~Wichman, ``Effect of oscillator phase noise
  and processing delay in full-duplex ofdm repeaters,'' in \emph{Proc. 46th
  Asilomar Conference on Signals, Systems and Computers}, Nov. 2012, pp.
  1947--1951.

\bibitem{Sahai12}
A.~Sahai, G.~Patel, C.~Dick, and A.~Sabharwal, ``Understanding the impact of
  phase noise on active cancellation in wireless full-duplex,'' in \emph{Proc.
  46th Asilomar Conference on Signals, Systems and Computers}, Nov. 2012, pp.
  29--33.

\bibitem{Syrjala13}
V.~Syrj\"{a}l\"{a}, M.~Valkama, L.~Anttila, T.~Riihonen, and D.~Korpi,
  ``Analysis of oscillator phase-noise effects on self-interference
  cancellation in full-duplex {OFDM} radio transceivers,'' \emph{IEEE
  Transactions on Wireless Communications}, 2014.

\bibitem{Sahai122}
A.~Sahai, G.~Patel, and A.~Sabharwal, ``Asynchronous full-duplex wireless,'' in
  \emph{Proc. Fourth International Conference on Communication Systems and
  Networks (COMSNETS)}, Jan. 2012, pp. 1--9.

\bibitem{Ahmed133}
E.~Ahmed, A.~Eltawil, and A.~Sabharwal, ``Rate gain region and design tradeoffs
  for full-duplex wireless communications,'' \emph{IEEE Transactions on
  Wireless Communications}, vol.~12, no.~7, pp. 3556--3565, Jul. 2013.

\bibitem{Riihonen122}
T.~Riihonen and R.~Wichman, ``Analog and digital self-interference cancellation
  in full-duplex {MIMO-OFDM} transceivers with limited resolution in {A/D}
  conversion,'' in \emph{Proc. 46th Asilomar Conference on Signals, Systems and
  Computers}, Nov. 2012, pp. 45--49.

\bibitem{Li11}
S.~Li and R.~Murch, ``Full-duplex wireless communication using transmitter
  output based echo cancellation,'' in \emph{Proc. Global Telecommunications
  Conference (GLOBECOM)}, Dec. 2011, pp. 1--5.

\bibitem{Bliss12}
D.~Bliss, T.~Hancock, and P.~Schniter, ``Hardware phenomenological effects on
  cochannel full-duplex {MIMO} relay performance,'' in \emph{Proc. 46th
  Asilomar Conference on Signals, Systems and Computers}, Nov. 2012, pp.
  34--39.

\bibitem{Zheng13}
G.~Zheng, I.~Krikidis, and B.~Ottersten, ``Full-duplex cooperative cognitive
  radio with transmit imperfections,'' \emph{IEEE Transactions on Wireless
  Communications}, vol.~12, no.~5, pp. 2498--2511, May 2013.

\bibitem{Korpi13}
D.~Korpi, T.~Riihonen, V.~Syrj\"{a}l\"{a}, L.~Anttila, M.~Valkama, and
  R.~Wichman, ``Full-duplex transceiver system calculations: Analysis of {ADC}
  and linearity challenges,'' \emph{IEEE Transactions on Wireless
  Communications}, 2014.

\bibitem{Anttila11}
L.~Anttila, ``Digital front-end signal processing with widely-linear signal
  models in radio devices,'' Ph.D. dissertation, Tampere University of
  Technology, 2011.

\bibitem{LTE_specs}
``{LTE}; evolved universal terrestrial radio access ({E-UTRA}); user equipment
  ({UE}) radio transmission and reception ({3GPP TS} 36.101 version 11.2.0
  release 11),'' ETSI, Sophia Antipolis Cedex, France.

\bibitem{Hua12}
Y.~Hua, P.~Liang, Y.~Ma, A.~Cirik, and Q.~Gao, ``A method for broadband
  full-duplex {MIMO} radio,'' \emph{Signal Processing Letters}, vol.~19,
  no.~12, pp. 793--796, Dec. 2012.

\bibitem{maxim_datasheet}
``{MAX2829} single-/dual-band 802.11a/b/g world-band transceiver {IC},'' Maxim
  Integrated, San Jose, California, USA.

\bibitem{Gu06}
Q.~Gu, \emph{{RF} System Design of Transceivers for Wireless
  Communications}.\hskip 1em plus 0.5em minus 0.4em\relax Springer-Verlag New
  York, Inc., 2006.

\bibitem{Parssinen99}
A.~P\"{a}rssinen, J.~Jussila, J.~Ryyn\"{a}nen, L.~Sumanen, and K.~A.~I.
  Halonen, ``A 2-{GHz} wide-band direct conversion receiver for {WCDMA}
  applications,'' \emph{IEEE Journal of Solid-State Circuits}, vol.~34, no.~12,
  pp. 1893--1903, Dec. 1999.

\bibitem{Yoshida03}
H.~Yoshida, T.~Kato, T.~Toyoda, I.~Seto, R.~Fujimoto, T.~Kimura, O.~Watanabe,
  T.~Arai, T.~Itakura, and H.~Tsurumi, ``Fully differential direct conversion
  receiver for {W-CDMA} using an active harmonic mixer,'' in \emph{Proc. Radio
  Frequency Integrated Circuits Symposium}, Jun. 2003, pp. 395--398.

\bibitem{Schreier10}
P.~Schreier and L.~Scharf, \emph{Statistical Signal Processing of
  Complex-Valued Data: The Theory of Improper and Noncircular Signals}.\hskip
  1em plus 0.5em minus 0.4em\relax Cambridge University Press, 2010.

\bibitem{Picinbono95}
B.~Picinbono and P.~Chevalier, ``Widely linear estimation with complex data,''
  \emph{IEEE Transactions on Signal Processing}, vol.~43, no.~8, pp.
  2030--2033, Aug. 1995.

\bibitem{Ding04}
L.~Ding, G.~Zhou, D.~Morgan, Z.~Ma, J.~Kenney, J.~Kim, and C.~Giardina, ``A
  robust digital baseband predistorter constructed using memory polynomials,''
  \emph{IEEE Transactions on Communications}, vol.~52, no.~1, pp. 159--165,
  Jan. 2004.

\bibitem{Morgan06}
D.~Morgan, Z.~Ma, J.~Kim, M.~Zierdt, and J.~Pastalan, ``A generalized memory
  polynomial model for digital predistortion of {RF} power amplifiers,''
  \emph{IEEE Transactions on Signal Processing}, vol.~54, no.~10, pp.
  3852--3860, Oct. 2006.

\bibitem{Hammi08}
O.~Hammi, F.~Ghannouchi, and B.~Vassilakis, ``On the sensitivity of {RF}
  transmitters' memory polynomial model identification to delay alignment
  resolution,'' \emph{Microwave and Wireless Components Letters}, vol.~18,
  no.~4, pp. 263--265, Apr. 2008.

\end{thebibliography}


\begin{IEEEbiography}[{\includegraphics[width=1in,height=1.25in,clip,keepaspectratio]{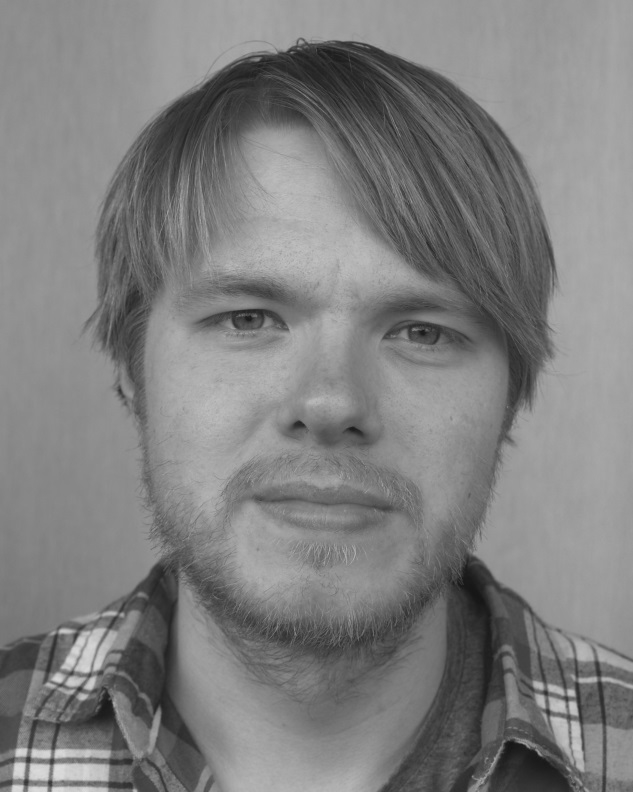}}]{Dani Korpi}
was born in Ilmajoki, Finland, on November 16, 1989. He received the B.Sc. and M.Sc. degrees (both with honours) in communications engineering from Tampere University of Technology (TUT), Finland, in 2012 and 2014, respectively.

He is currently a researcher at the Department of Electronics and Communications Engineering at TUT, pursuing the D.Sc. (Tech.) degree in communications engineering. His main research interest is the study and development of single-channel full-duplex radios, with a focus on analysing the RF impairments.
\end{IEEEbiography}

\begin{IEEEbiography}[{\includegraphics[width=1in,height=1.25in,clip,keepaspectratio]{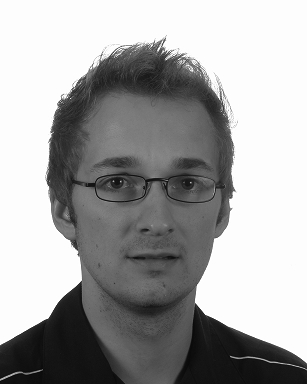}}]{Lauri Anttila}
(S'06, M'11) received the M.Sc. (Tech.) degree in 2004 and the D.Sc. (Tech.) degree (with honours) in 2011 in communications engineering from Tampere University of Technology (TUT), Tampere, Finland.

Currently, he is a Research Fellow at the Department of Electronics and Communications Engineering at TUT. His general research interests include statistical and adaptive signal processing for communications, digital front-end signal processing in flexible radio transceivers, radio architectures, and full-duplex radio systems. 
\end{IEEEbiography}

\begin{IEEEbiography}[{\includegraphics[width=1in,height=1.25in,clip,keepaspectratio]{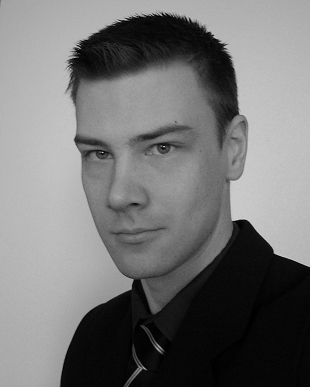}}]{Ville Syrj\"al\"a}
(S'09, M'12) was born in Lapua, Finland, in 1982. He received the M.Sc. (Tech.) degree in 2007 and D.Sc. (Tech.) degree in 2012 in communications engineering (CS/EE) from Tampere University of Technology (TUT), Finland.

He was working as a research fellow with the Department of Electronics and Communications Engineering at TUT, Finland, until 2013. Currently, he is working as a research fellow of the Japan Society for the Promotion of Science (JSPS) at Kyoto University, Japan. His general research interests are in full-duplex radio technology, communications signal processing, transceiver impairments, signal processing algorithms for flexible radios, transceiver architectures, direct sampling radios, and multicarrier modulation techniques.
\end{IEEEbiography}

\begin{IEEEbiography}[{\includegraphics[width=1in,height=1.25in,clip,keepaspectratio]{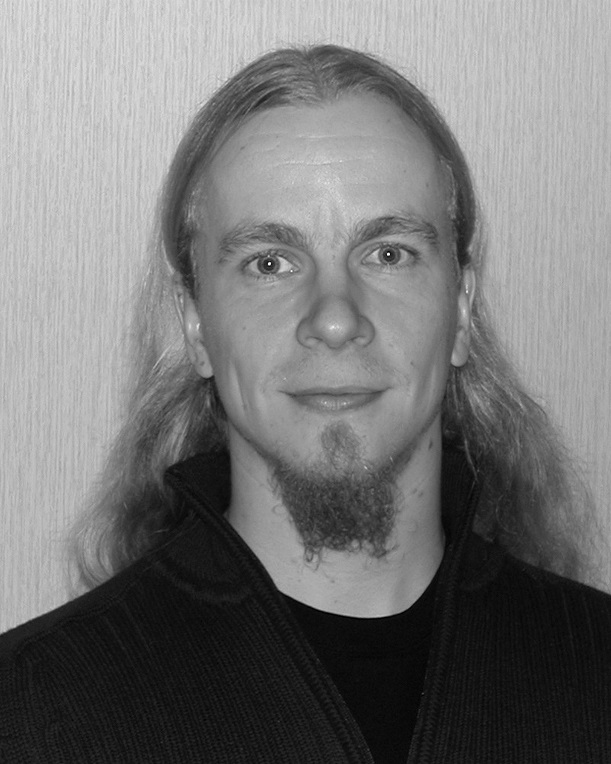}}]{Mikko Valkama}
(S'00, M'02) was born in Pirkkala, Finland, on November 27, 1975. He received the M.Sc. and Ph.D. degrees (both with honours) in electrical engineering (EE) from Tampere University of Technology (TUT), Finland, in 2000 and 2001, respectively. In 2002 he received the Best Ph.D. Thesis award by the Finnish Academy of Science and Letters for his dissertation entitled "Advanced I/Q signal processing for wideband receivers: Models and algorithms".

In 2003, he was working as a visiting researcher with the Communications Systems and Signal Processing Institute at SDSU, San Diego, CA. Currently, he is a Full Professor and Department Vice Head at the Department of Electronics and Communications Engineering at TUT, Finland. He has been involved in organizing conferences, like the IEEE SPAWC'07 (Publications Chair) held in Helsinki, Finland. His general research interests include communications signal processing, estimation and detection techniques, signal processing algorithms for software defined flexible radios, full-duplex radio technology, cognitive radio, digital transmission techniques such as different variants of multicarrier modulation methods and OFDM, radio localization methods, and radio resource management for ad-hoc and mobile networks.
\end{IEEEbiography}

\end{document}